\documentclass{llncs}
   \usepackage{times}
   \usepackage{mathptm}
   \usepackage{epsfig}
   \usepackage{color}
   \usepackage{latexsym}
   \usepackage{pslatex}
   \usepackage{cite}
   \usepackage{graphicx}
   \usepackage{algorithm}
   \usepackage{algorithmic}
   \usepackage{listings}
   \usepackage{multirow}
   \usepackage{pslatex}
   \usepackage{tabularx}
   \usepackage{colortbl}
   \usepackage{amsfonts}

\newenvironment{keywords}{
       \list{}{\advance\topsep by0.35cm\relax\small
       \leftmargin=1cm
       \labelwidth=0.35cm
       \listparindent=0.35cm
       \itemindent\listparindent
       \rightmargin\leftmargin}\item[\hskip\labelsep
                                     \bfseries Keywords:]}
     {\endlist}

\newtheorem{defn}{Definition}
\newtheorem{prop}{Property}

\newcommand{\procname}[1]{\textsc{#1} }
\newcommand{\ALGORITHM}{\textbf{algorithm} }

\newcommand{\ELSE}{\textbf{else} }

\newcommand{\END}{\textbf{end} }

\newcommand{\DO}{\textbf{do} }

\newcommand{\FOR}{\textbf{for} }
\newcommand{\FOREACH}{\textbf{for each} }
\newcommand{\IF}{\textbf{if} }

\newcommand{\RETURN}{\textbf{return} }

\newcommand{\THEN}{\textbf{then} }

\newcommand{\WHILE}{\textbf{while} }

\newcommand{\BREAK}{\textbf{break} }

\newcommand{\INPUT}{\textbf{input:} }

\begin{document}
\thispagestyle{plain}
\pagestyle{plain}  

\title{A Linear-Time Algorithm for Finding All Double-Vertex Dominators of a Given Vertex}
\author{Maxim~Teslenko\inst{1} \and Elena Dubrova\inst{2}}
\institute{Ericsson Research, Ericsson, F\"ar\"ogatan 6, 164 80 Stockholm, Sweden\\
\email{maxim.teslenko@ericsson.com}
\and
Royal Institute of Technology, Electrum 229, 164 40 Stockholm, Sweden\\
\email{dubrova@kth.se}
}

\maketitle
 
\begin{abstract}

Dominators provide a general mechanism for identifying reconverging paths in graphs. This is useful for a number of applications in Computer-Aided Design (CAD) including signal probability computation in biased random simulation, switching activity estimation in power and noise analysis, and cut points identification in equivalence checking. However, traditional single-vertex dominators are too rare in circuit graphs. In order to handle reconverging paths more efficiently, we consider the case of double-vertex dominators which occur more frequently. First, we derive a number of specific properties of double-vertex dominators. Then, we describe a data structure for representing all double-vertex dominators of a given vertex in linear space. Finally, we present an algorithm for finding all double-vertex dominators of a given vertex in linear time. Our results provide an efficient systematic way of partitioning large graphs along the reconverging points of the signal flow.
\end{abstract} 

\begin{keywords} 
Graph, dominator, min-cut, logic circuit, reconverging path
\end{keywords} 
\section{Introduction}
\label{ddom_sec:introduction}

\begin{figure}[t!]
\begin{center}
\includegraphics*[width=4in]{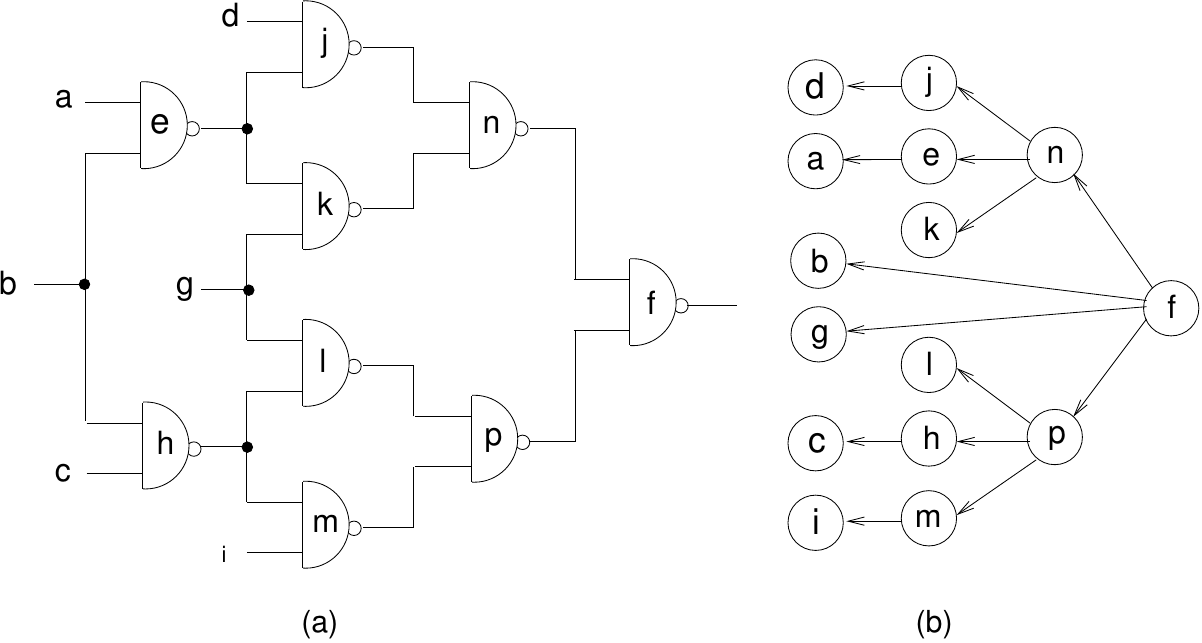}
\caption{(a) An example circuit; (b) Its dominator tree.}
\label{ddom_f2}
\end{center}
\end{figure}

This paper considers the problem of finding dominators in circuit graphs. 
A vertex $v$ is said to {\em dominate} another vertex $u$ if every path
from $u$ to the output of the circuit contains $v$~\cite{LeT79}.  
For example, for the circuit in Figure~\ref{ddom_f2}(a), vertex $n$
dominates vertex $e$; vertex $p$ dominates vertex $h$, etc.

Dominators provide a general mechanism for identifying re-converging
paths in graphs. If a vertex $v$ is the origin of a
re-converging path, then the immediate dominator of $v$ is the
earliest point at which such a path converges. 
For example, in
Figure~\ref{ddom_f2}(a), the re-converging path originated at $e$ ends at $n$;
the re-converging path originated at $g$ ends at $f$.

Knowing the precise starting and ending points of a re-converging path
is useful in a number of applications including computation of signal
probabilities in biased random simulation, estimation of switching activities in power and
noise analysis, and identification of cut points in
equivalence checking.

The {\em signal probability} of a net in a combinational circuit is
the probability that a randomly generated input vector will produce
the value one on this net~\cite{PaM75}. Signal probability
analysis is used, for example, to measure and control the coverage of vector
generation for biased random simulation~\cite{KuLWCH07}.

The average {\em switching activity} in a
combinational circuit is the probability of its net values to change
from 0 to 1 or vice verse~\cite{GhDK92}.  It correlates directly with the average
dynamic power dissipation of the circuit, thus its analysis is
useful for guiding logic optimization methods targeting low power consumption~\cite{CoMD97}.

Computation of signal probabilities and switching activities based on
topologically processing the circuit from inputs to outputs and
evaluating the gate functions generally produces incorrect results due
to higher-order exponents introduced by correlated signals~\cite{PaM75}. For
example, if the functions $f$ and $g$ have variables in common, then
$P[f\wedge g] \not= P[f]\cdot P[g]$, where $P$ is the signal
probability.  Dominators provide the earliest points during
topological processing at which all signals correlated with signal originated at the dominated vertex
converge. Therefore, the computation of
signal probabilities and switching activities can be 
partitioned along the dominator points.

Cut-points
based equivalence checking partitions the specification and implementation 
circuits along frontiers of functionally equivalent signal pairs, called 
{\em cut-points}~\cite{KhMKH01}.  This is usually done in four steps: (1) cut-points
identification, attempting to discover as many cut-points as possible,
(2) cut-points selection, aiming to choose the cut-points which
simplify the task of verification, (3) equivalence checking of the
resulting sub-circuits, (4) false negative reduction.
Dominators provide a systematic mechanism for identifying and choosing good cut-points
in circuits, since converging points of the signal flow are ideal candidates for cut-points.

In spite of the theoretical advantages of dominators,
previous attempts to apply dominator-based techniques to large circuits
have not been successful. 
Two main reasons for this are: (1) single-vertex dominators,
which can be found in linear time, are too rare in circuits; (2)
multiple-vertex dominators, which are common in circuits, require
exponential time to be computed. 
In other words, no systematic approach
for finding useful dominators in large circuits efficiently has been known so far.
Useful are normally dominators of a
small size because $2^k$ combinations of values
of a $k$-vertex dominator have to be manipulated to resolve signal correlations~\cite{KrDK03}.

In this paper, we focus on the specific case double-vertex dominators.  
First, we prove a number of fundamental properties of double-vertex dominators.
For example, we show that immediate double-vertex dominators are unique.
This property also holds for single-vertex dominators, but it does not extend
to dominators of size larger than two. Then, we present a data structure for representing
all double-vertex dominators of a given vertex in linear space. 
Finally, we introduce an algorithm for finding
all double-vertex dominators of a given vertex in linear time.
This asymptotically reduces the complexity of the previous quadratic
algorithm for finding double-vertex dominators~\cite{TeD05b}.


The paper is organized as follows. Section~\ref{basic} 
presents basic notation and definitions.
In Section~\ref{ddom_prelim}, we introduce
definitions of dominators which are more
general than the traditional ones from~\cite{LeT79}. 
Section~\ref{ddom_prev} summarizes the
previous work on dominators.  
In Sections~\ref{ddom_mdp} and~\ref{sec_dd}, we describe properties of 
multiple-vertex and double-vertex dominators, respectively.
Section~\ref{ddom_ds} presents the data
structure for representing double-vertex dominators.
Section~\ref{ddom_alg} describes the new algorithm for finding double-vertex dominators. 
The experimental results are shown in
Section~\ref{ddom_exp}. Section~\ref{ddom_con} concludes the paper.

\section{Preliminaries} \label{basic}

Unless otherwise specified, throughout the paper, we use capital letters $A, B, C,$ etc.
to denote vectors and bold letters $\mathbb{A}, \mathbb{B}, \mathbb{C},$ etc. to denote sets.

Let $G = (\mathbb{V},\mathbb{E},root)$ denote a single-output acyclic circuit graph where the set
of vertices $\mathbb{V}$ represents the primary inputs and gates.  A particular
vertex $root \in \mathbb{V}$ is marked as the circuit output. The set of edges
$\mathbb{E}\subseteq \mathbb{V} \times \mathbb{V}$ represents the nets connecting the gates.

{\em Fanin} and {\em fanout} sets of a vertex $v \in
\mathbb{V}$ are defined as $fanin(v) = \{u \ | (u,v) \in E\}$ and $fanout(v) =
\{u \ | (v,u) \in E\}$, respectively.

The {\em transitive fanin} of a vertex $v \in \mathbb{V}$ is a subset of
$\mathbb{V}$ containing all vertices from which $v$ in reachable.
Similarly, the {\em transitive fanout} of a vertex $v \in \mathbb{V}$ is a subset of
$\mathbb{V}$ containing all vertices reachable from $v$.

A {\em path} $P = (v_1, v_2, \ldots, v_{|P|})$ is a vector of vertices of $\mathbb{V}$
such that $(v_i,v_{i+1}) \in \mathbb{E}$ for all $i \in \{1, \ldots, |P|-1\}$. The
vertices $v_1$ and $v_{|P|}$ are called the {\em source} and the {\em
sink} of $P$, respectively. The source and the sink of $P$ are
called the {\em terminal} vertices of $P$. The remaining vertices of 
$P$ are called the {\em non-terminal} vertices. 

Throughout the paper, we call two paths {\em disjoint} if the intersection of
sets of their non-terminal vertices is empty.

Given two paths $P_1=(v_1, v_2, \ldots, v_{|P_1|})$ and $P_2=(w_1, w_2,
\ldots, w_{|P_2|})$, the {\em concatenation} of $P_1$ and $P_2$
is defined only if $v_{|P_1|} = w_1$. The result of the
concatenation is the path $P_3 = (v_1, v_2, \ldots, v_{|P_1|}, w_2, \ldots,
w_{|P_2|})$.  We use the notation $P_3 = P_1 P_2$ to denote that $P_3$
is a concatenation of $P_1$ and $P_2$.

A {\em prefix} of a vertex $P$, denoted by {\em prefix}$(P)$, is a sub-vertex of 
$P$ containing $k$ first adjacent vertices of $P$ for some $1 \leq k < |P|$. 
A {\em suffix} of a vertex $P$, denoted by {\em suffix}$(P)$,
is a sub-vertex of $P$ containing $k$ last adjacent vertices of $P$ for some $1 < k \leq |P|$. 

\section{Definition of Dominators} \label{ddom_prelim}

In this section, we introduce definitions of dominators and immediate
dominators which are more general than the traditional ones from~\cite{LeT79}.

\begin{defn} \label{ddom_dom}
A set of vertices $\mathbb{A}$ {\em dominates} a set of vertices $\mathbb{B}$ with
respect to a set of vertices $\mathbb{C}$ if every path which starts at a vertex
in $\mathbb{B}$ and ends at a vertex in $\mathbb{C}$ contains at least one vertex from
$\mathbb{A}$. 
\end{defn}

\begin{defn} \label{ddom_dom2}
A set of vertices $\mathbb{A}$ is a {\em dominator} of a set of vertices
$\mathbb{B}$ with respect to a set of vertices $\mathbb{C}$, if
\begin{itemize}
\item[(a)] $\mathbb{A}$ dominates $\mathbb{B}$,
\item[(b)] $\forall v \in \mathbb{A}$, $\mathbb{A}-\{v\}$ does not dominate $\mathbb{B}$.
\end{itemize}
\end{defn}

The sets $\mathbb{B}$ and $\mathbb{C}$ are called, the {\em source} set and the {\em sink} set, respectively.
For example, for the circuit in Figure~ \ref{ddom_f2}(a), $\{j,k,l\}$ is a dominator of the source $\{e,g\}$
with respect to the sink $\{n,p\}$.

In most applications of dominators, the source set $\mathbb{B}$ and the
sink set $\mathbb{C}$ are known, while the dominator set $\mathbb{A}$ needs to
be computed. The sizes of the sets $\mathbb{B}$,
$\mathbb{C}$ are neither important for the choice of data structure for representing dominators,
nor for the algorithm which finds them. Vertices in the set $\mathbb{B}$
can be merged into a single vertex $v_b$ which feeds all the vertices fed by any vertex in $\mathbb{B}$. Similarly, vertices into the set $\mathbb{C}$
can be merged to a single vertex $v_c$ which is fed by all vertices feeding any vertex in $\mathbb{C}$. In this case finding a dominator for $v_b$ with respect to $v_c$ is equivalent to finding a dominator for $\mathbb{B}$ with respect to $\mathbb{C}$. Therefore, an algorithm which
handles the case $|\mathbb{B}|=|\mathbb{C}|=1$ can be extended to the sets $\mathbb{B}$ and $\mathbb{C}$ of an arbitrary size.

Contrary, the size of the dominator set $\mathbb{A}$ is crucial for the choice
of data structures and algorithms.  
Therefore, the size of $\mathbb{A}$
is the most important criteria for characterizing the properties of a
dominator. 
We use the term $k$-{\em vertex dominator} to refer to
the case of $|\mathbb{A}| = k$. If $k>1$ then we may also call a $k$-vertex dominator {\em multiple-vertex dominator}.
If a dominator dominates more then one vertex, i.e. $|B|>1$, it is called
{\em common} $k$-vertex dominator.

Throughout this paper, unless specified otherwise, the vertex $root$
is assumed to be the sink for any considered dominator relation.
So, if we say that $\mathbb{A}$ dominates $\mathbb{B}$, we mean that $\mathbb{A}$ dominates
$\mathbb{B}$ with respect to $root$.

\begin{defn} \label{ddom_sdom}
A set of vertices $\mathbb{A}$ is a {\em strict} dominator of a set of vertices $\mathbb{B}$,
if $\mathbb{A}$ is a dominator of $\mathbb{B}$ and $A \bigcap B = \emptyset$.
\end{defn}

For example, in Figure~ \ref{ddom_f2}(a), $\{j,k,h\}$ is a dominator of $\{b,h\}$,
but it is not strict. On the other hand, $\{j,k,h\}$ is a strict dominator of $\{b\}$. Obviously, any
dominator of a single vertex is a strict dominator. All results in this paper are derived for 
dominators of single vertices. Therefore, throughout the paper when we write "dominator" it also means "strict dominator".
Note that 
any algorithm which finds only strict dominators can be 
extended to find all dominators by introducing a
fake vertex which feeds all nodes in $\mathbb{B}$. The search is
carried out with the fake vertex constituting the new $\mathbb{B}$.

\begin{defn} \label{ddom_idom}
A set $\mathbb{A}$ is an {\em immediate} $k$-vertex dominator of a set $\mathbb{B}$ if $\mathbb{A}$ is
a strict $k$-vertex dominator of $\mathbb{B}$ and $\mathbb{A}$ does not dominate $\mathbb{D}$, where $\mathbb{D}$ is
any other strict $k$-vertex dominator of $\mathbb{B}$.
\end{defn}

The concept of immediate dominators has a special importance for
single-vertex dominators. It was shown
in~\cite{LoM69,AhU72} that every vertex $v$ in a directed acyclic graph $G$ except $root$ has a
unique immediate single-vertex dominator, $idom(v)$.
The edges $\{(idom(v),v) \ | \ v \in \mathbb{V} - \{root\}\}$ form a directed
tree rooted at $root$, which is called the {\em dominator tree} of
$G$. For example, the dominator tree for the circuit  in Figure~\ref{ddom_f2}(a) is shown in Figure~\ref{ddom_f2}(b).

Note that the immediate multiple-vertex dominators are not necessarily unique.
For example, vertex $b$ in Figure~\ref{ddom_f2}(a) has two immediate 3-vertex dominators: $\{j,k,h\}$ and $\{e,l,m\}$.
Later in the paper we prove that the immediate dominators 
are always unique for the case of $k = 2$.

It might be worth mentioning that dominators are more general than
{\em min-cut} in circuit partitioning~\cite{KeL70}. A min-cut is
required to dominate all vertices in its transitive fanin. Therefore, every
min-cut is a dominator, but not every dominator is a min-cut.

\section{Previous Work} \label{ddom_prev}

The problem of finding single-vertex dominators was first considered
in global flow analysis and program optimization. Lorry and
Medlock~\cite{LoM69} presented an $O(n^4)$ algorithm for finding
all immediate single-vertex dominators in
a flowgraph with $n$ vertices.
Successive improvements of this algorithm were done
by Aho and Ullman~\cite{AhU72}, Purdom and Moore~\cite{PuM72}, and
Tarjan~\cite{Ta74}, culminating in Lengauer and Tarjan's~\cite{LeT79}
$O(e \alpha(e,n))$ algorithm, where $e$ is the number of edges and
$\alpha$ is the standard functional inverse of the Ackermann function
which grows slowly with $e$ and $n$.

The asymptotic time complexity of finding single-vertex dominators was
reduced to linear by Harel~\cite{Ha85}, Alstrup et al.~\cite{AlHLT99}
and Buchsbaum et al.~\cite{BuKRW98}. However, these improvements in
asymptotic complexity did not contribute much to reducing the actual
runtime. For example, the algorithm~\cite{BuKRW98} runs 10\% to 20\%
slower than Lengauer and Tarjan's~\cite{LeT79}. Lengauer and Tarjan
algorithm appears to be the fastest of algorithms for single-vertex
dominators on graphs of large size.

One of the first attempts to develop an algorithm for the
identification of multiple-vertex dominators was done by
Gupta. In~\cite{gupta}, three algorithms addressing this
problem were proposed. The first finds all immediate multiple-vertex dominators of
size up to $k$ in $O(n^k)$ time. Computing immediate dominators is
easy because an immediate dominator of a vertex $v$ is always
contained in the set of fanout vertices of $v$. Possible redundancies
can be removed by checking whether for every $u$ in the fanout of
$v$ there exists at least one path from $u$ to $root$ which contains
$u$ and does not contain any other $w$ in the fanout of $v$.

The second algorithm in~\cite{gupta} finds all multiple-vertex dominators of
a given vertex. The number of all dominators of a vertex can be
exponential with respect to $n$. Since the algorithm represents each
dominator explicitly as a set of vertices, it has exponential space
and time complexity.

The third algorithm in~\cite{gupta} finds all multiple-vertex dominators of
size up to $k$ for all vertices in the circuit. Due to its specific
nature, this algorithm cannot not be modified to search for all
multiple-vertex dominators of a fixed size for a given vertex.
The complexity of the algorithm is not evaluated in the
paper. Depending on the implementation, the complexity can vary from
exponential to polynomial with a high degree of the polynomial.
For example, for double-vertex dominators, the complexity of the
algorithm is at least $O(n^5)$.

Successive improvements of the algorithms in~\cite{gupta} were done in~\cite{KrD05a,KrD05b,Du04} and~\cite{DuTM04}.
The algorithm presented in~\cite{DuTM04} finds the set of all
possible $k$-vertex dominators of a circuit by
iteratively restricting the graph with respect to one of its vertices, $v$. The restriction is done by removing from 
the graph all vertices dominated by $v$. Dominators of size $k-1$ are then computed for the resulting
restricted graph by applying the same
technique recursively.  Once $k$ is reduced to 1, a single-vertex
dominator algorithm is used.  Since single-vertex dominators can be
computed in linear time, the overall complexity of the
algorithm~\cite{DuTM04} is bounded by $O(n^k)$. 

The first algorithm designed specifically for
double-vertex dominator was presented in~\cite{TeD05b}. This algorithm
uses the max-flow algorithm to find an immediate double-vertex dominator 
for a given set of vertices $\mathbb{B}$. The immediate dominator is
considered as a sink and all vertices in $\mathbb{B}$ are merged into a single
source vertex.  The obtained min-cut corresponds to the minimal-size 
dominator which dominates all paths from the source to the sink.  
If the size of the min-cut is larger than two, then $S$ does not
have any double-vertex dominators. The complexity of this algorithm 
is $O(n^2)$.

Interesting results on testing two-connectivity of directed graphs in linear time were presented in~\cite{AbGKM10}, with a focus on finding disjoint paths. Since dominators are contained in disjoint
paths, the results of~\cite{AbGKM10} can potentially facilitate their search. However, with such an approach, the complexity of checking if a pair of vertices is a double-vertex dominator remains
linear. As we show later, in our case it is reduced to a constant.

The {\em cactus tree} data structure for representing all undirected min-cuts was introduced
in~\cite{DiKL76}. The problem of finding a min-cut of a high degree is reduced to finding a two-element
cut in the cactus tree. Such a structure allows for extracting min-cuts of a high degree, which
are a special case of $k$-vertex dominators. In our case, the original degree is two. Therefore,
the cactus tree data structure cannot help reduce is further.

\section{Properties of Multiple-Vertex Dominators} \label{ddom_mdp}

In this section, we derive some general properties of $k$-vertex
dominators. 
The following three Lemmata show antisymmetry, transitivity, and reflexivity
of the dominator relation.

\begin{lemma} \label{ddom_mp01}
Let $\mathbb{A}$ and $\mathbb{B}$ be two different dominators of a vertex $u$. If $\mathbb{B}$
dominates $\mathbb{A}$, then $\mathbb{A}$ does not dominate $\mathbb{B}$.
\end{lemma}
{\bf Proof:} Set $\mathbb{A}$ is not equal to $\mathbb{B}$ by the condition of the Lemma. $\mathbb{A}$
is not a proper subset of $\mathbb{B}$ either, because otherwise $\mathbb{B}$ would
violate the Definition~\ref{ddom_dom2}b. Thus, there is a
vertex $v \in A$ such that $v \not \in B$.  Since $\mathbb{A}$ is a dominator of $u$, by
Definition~\ref{ddom_dom2}b, there exists $P=(u, \ldots, root)$, such that $v \in
P$, and $v_2 \not \in P$, $\forall v_2 \in (A-\{v\})$.  The path $P_2=(v,
\ldots, root)$ which is suffix of $P$ should contain a vertex $w \in
B$ since $\mathbb{B}$ dominates $\mathbb{A}$. The path $P_3=(w, \ldots, root)$ which is
a suffix of $P_2$ does not contain any vertex of $\mathbb{A}$ by
construction. Thus, by Definition~\ref{ddom_dom}, $\mathbb{B}$ does not dominate
$\mathbb{A}$. 
\begin{flushright}
$\Box$ \\
\end{flushright}

\begin{lemma} \label{ddom_mp02}
If $\mathbb{A}$ dominates $\mathbb{B}$ and $\mathbb{B}$ dominates $\mathbb{C}$, then $\mathbb{A}$ dominates $\mathbb{C}$.
\end{lemma}
{\bf Proof:} Consider an arbitrary path $P=(v, \ldots, root)$ such that $v
\in \mathbb{C}$. We proof the Lemma by showing that a vertex from $\mathbb{A}$ is in
$P$.  Since $\mathbb{B}$ dominates $\mathbb{C}$, it holds that $\exists w \in B$ such
that $w \in P$.  The path $P_2=(w, \ldots, root)$ is a suffix of $P$.
Since $\mathbb{A}$ dominates $\mathbb{B}$, it holds that $\exists u \in A$ such that $u
\in P_2$.  Thus $u \in P$ as well. 
\begin{flushright}
$\Box$ \\
\end{flushright}

\begin{lemma} \label{ddom_mp03}
$\mathbb{A}$ dominates $\mathbb{A}$.
\end{lemma}
{\bf Proof:}
Follows trivially from the Definition~\ref{ddom_dom}a. 
\begin{flushright}
$\Box$ \\
\end{flushright}

It follows from the above three Lemmata that any set of dominators of
a vertex $u$ is partially ordered by the dominator relation.


\section{Properties of Double-Vertex Dominators} \label{sec_dd}
\label{ddom_ddp}

In this section, we derive a number of fundamental properties of
double-vertex dominators. 


Let $\mathbb{D}_u$ be the set of all possible double-vertex dominators of a vertex $u \in \mathbb{V}$. 
Each element of $\mathbb{D}_u$ is a pair of vertices $\{v,w\}$, $v,u \in \mathbb{V}$, constituting a double-vertex dominator of $u$. 
With some abuse of notation, throughout the paper we write $v \in \mathbb{D}_u$ as a shorthand for
$\exists w \in \mathbb{V}$ such that $\{v,w\} \in \mathbb{D}_u$.

The following Lemma shows that if two dominators have a common vertex, 
then one of the dominators dominates the non-common vertex in another dominator.

\begin{lemma} \label{ddom_p01}
If $\{v_1,v_2\} \in \mathbb{D}_u$ and $\{v_2,v_3\} \in \mathbb{D}_u$,
then either $\{v_1,v_2\}$ dominates $v_3$, or $\{v_2,v_3\}$ dominates $v_1$.
\end{lemma}
{\bf Proof:} If $\{v_1,v_2\}$ dominates $v_3$, then the Lemma
holds trivially. Suppose that $\{v_1,v_2\}$ does not dominate $v_3$. Since
$\{v_2,v_3\} \in \mathbb{D}_u$, by Definition~\ref{ddom_dom2}b, there exists $P_1=(u,
\ldots, root)$, such that $v_3 \in P_1$ and $v_2 \not \in P_1$. Since
$\{v_1,v_2\} \in \mathbb{D}_u$, for all   $P_1$ it holds that $v_1 \in
P_1$. Furthermore, $v_1$ precedes $v_3$ in $P_1$, because, by
assumption, $\{v_1,v_2\}$ does not dominate $v_3$. Thus the prefix $P_2=(u,
\ldots, v_1)$ of the path $P_1$ does not contain $v_2$ and $v_3$.

Then, there exists no path $P_3=(v_1, \ldots, root)$ such that $v_2,v_3 \not
\in P_3$, because otherwise the path $P_2 P_3$ would contain neither
$v_2$ nor $v_3$. This would contradict $\{v_2,v_3\} \in \mathbb{D}_u$. So
for all $P_3$, it holds that either $v_2 \in P_3$ or $v_3 \in
P_3$. Thus, by Definition~\ref{ddom_dom}, $\{v_2,v_3\}$ dominates $v_1$.

Similarly we can show that if $\{v_2,v_3\}$ does not dominate $v_1$, then
$\{v_1,v_2\}$ dominates $v_3$.
\begin{flushright}
$\Box$
\end{flushright}

The following Lemma considers the case of two double-vertex dominators
which have no vertices in common and which do not dominate each other.

\begin{lemma} \label{ddom_p03}
If $\{v_1,v_2\} \in \mathbb{D}_u$, $\{v_3,v_4\} \in \mathbb{D}_u$,
$\{v_3,v_4\}$ does not dominate $v_1$, and $\{v_1,v_2\}$ does not
dominate $v_4$, then $\{v_1,v_4\} \in \mathbb{D}_u$ and $\{v_2,v_3\} \in
\mathbb{D}_u$.

\end{lemma}
{\bf Proof:} Vertices $v_1,v_2,v_3$ and $v_4$ belong to $\mathbb{D}_u$. Thus,
 none of them is a single-vertex dominator of
$u$. Therefore, any deduction showing that any pair of these vertices
dominates $u$ would imply that this pair is a double-vertex
dominator of $u$.

First, we show that $\{v_2,v_3\} \in \mathbb{D}_u$. 
Consider the following two cases:
\begin{itemize}
\item[(1)] There exists $P_3=(u, \ldots, root)$ such that $v_1,v_4 \in P_3$,
\item[(2)] There exists no $P_3=(u, \ldots, root)$ such that $v_1,v_4 \in P_3$.
\end{itemize}

\noindent {\bf Case 1:} One of the vertices $v_1$, $v_4$ precedes another one in $P_3$.

\noindent (a)  Assume that $v_1$ precedes $v_4$. This implies that
for all $P_4=(u, \ldots, v_1)$, $v_4 \not \in P_4$. According to
the conditions of the Lemma, $\{v_3,v_4\}$ does not dominate $v_1$.
This means that there exists $P_1={(v_1 , \ldots, root)}$ such that
$v_3,v_4 \not\in P_1$. Then, there exists no $P_4$ such that $v_3 \not
\in P_4$, or otherwise a path $P_4P_1$ would contain neither $v_3$ nor
$v_4$, and that would contradict $\{v_3,v_4\} \in \mathbb{D}_u$. 
So, for all $P_4$, $v_3 \in P_4$. Thus, every path $(u, \ldots, root)$
containing $v_1$ contains $v_3$ as well. Thus, $v_1$ can be substituted
by $v_3$ in any dominator of $u$. So $\{v_1,v_2\} \in \mathbb{D}_u$ implies
that $\{v_2,v_3\} \in \mathbb{D}_u$.

\noindent (b)  If $v_4$ precedes $v_1$, then the prove is similar to
(a) case. We can show that all paths $(u, \ldots, root)$ containing
$v_4$ contain $v_2$ as well. Thus, $\{v_3,v_4\} \in \mathbb{D}_u$ implies
that $\{v_2,v_3\} \in \mathbb{D}_u$.

\noindent {\bf Case 2:}
The assumption of the case 2 directly implies that for all
$P_4=(u, \ldots, v_1)$, $v_4 \not \in P_4$. The rest of the proof is
similar to the case 1(a).

Next, we show that $\{v_1,v_4\} \in \mathbb{D}_u$. Consider
two following two cases:
\begin{itemize}
\item[(1)] There exists $P_3=(u, \ldots, root)$ such that $v_2,v_3 \in P_3$,
\item[(2)] There exists no $P_3=(u, \ldots, root)$ such that $v_2,v_3 \in P_3$.
\end{itemize}

\noindent {\bf Case 1:} (a) Assume that $v_2$ precedes $v_3$. It
implies that, for all $P_4=(v_3, \ldots, root)$, $v_2
\not \in P_4$.  But $\{v_1,v_2\} \in Dom(v_3)$ implies that for all
$P_4$, $v_1 \in P_4$, i.e.  $v_1$ is a single-vertex dominator of
$v_3$.  Thus, $\{v_3,v_4\} \in \mathbb{D}_u$ implies that $\{v_1,v_4\} \in
\mathbb{D}_u$.

\noindent (b) If $v_3$ precedes $v_2$, the prove is similar to (a).
Then, $v_4$ is a single-vertex dominator of $v_2$.  Thus, $\{v_1,v_2\}
\in \mathbb{D}_u$ implies that $\{v_1,v_4\} \in \mathbb{D}_u$.

\noindent {\bf Case 2:}
The assumption of the case 2 directly implies that for all $P_4=(u
\ldots v_1)$, $v_4 \not \in P_4$. Thus $v_1$ is a single-vertex
dominator of $v_3$. Consequently $\{v_3,v_4\} \in \mathbb{D}_u$ implies that
$\{v_1,v_4\} \in \mathbb{D}_u$.
\begin{flushright}
$\Box$
\end{flushright}

The following Lemma shows another property of two double-vertex dominators
which have no vertices in common and which do not dominate each other.

\begin{lemma} \label{ddom_p04}
If $\{v_1,v_2\} \in \mathbb{D}_u$, $\{v_3,v_4\} \in \mathbb{D}_u$,
$\{v_3,v_4\}$ does not dominate $v_1$, and $\{v_1,v_2\}$ does not
dominate $v_4$, then $\{v_3,v_4\}$ dominates $v_2$ and $\{v_1,v_2\}$
dominates $v_3$.

\end{lemma}
{\bf Proof:} According to the Lemma~\ref{ddom_p03}, $\{v_2,v_3\} \in \mathbb{D}_u$
and $\{v_1,v_4\} \in \mathbb{D}_u$.

First, we prove that $\{v_2,v_3\}$ does not dominate $v_1$ by contradiction.
Assume that $\{v_2,v_3\}$ dominates $v_1$.

Since $\{v_1,v_2\} \in \mathbb{D}_u$, by Definition~\ref{ddom_dom2}b, there exists
$P_1 = (v_1, \ldots, root)$ such that $v_2 \not \in P_1$. Since
$\{v_2,v_3\}$ dominates $v_1$, this implies that $v_3 \in P_1$. Thus,
$v_1$ precedes $v_3$ in any path containing $v_1$, $v_3$.

Since $\{v_3,v_4\}$ does not dominate $v_1$, by Definition~\ref{ddom_dom},
there exists $P_2=( v_1, \ldots, root)$ such that $v_3 \not \in P_2$ and
$v_4 \not \in P_2$.

Since $\{v_1,v_4\} \in \mathbb{D}_u$, by Definition~\ref{ddom_dom2}b, there exists
$P_3= ( u, \ldots, v_1)$ such that $v_4 \not \in P_3$. Since $v_1$
precedes $v_3$, it implies that $v_3 \not \in P_3$ either.

The existence of the path $P_2P_3$ which does not contain neither $v_3$
nor $v_4$ contradicts the fact that $\{v_3,v_4\} \in \mathbb{D}_u$. Thus,
the assumption that $\{v_2,v_3\}$ dominates $v_1$ is invalid.

Since $\{v_1,v_2\} \in \mathbb{D}_u$ and $\{v_2,v_3\} \in \mathbb{D}_u$, according
to the Lemma~\ref{ddom_p01} either $\{v_1,v_2\}$ dominates $v_3$, or
$\{v_2,v_3\}$ dominates $v_1$. But, as we showed before, $\{v_2,v_3\}$
does not dominate $v_1$, thus $\{v_1,v_2\}$ dominates $v_3$.

The case of $\{v_3,v_4\}$ dominating $v_2$ can be proved similarly.
\begin{flushright}
$\Box$ \\
\end{flushright}

The following three Lemma consider mutual relations between of several dominators of the same vertex.

\begin{lemma} \label{ddom_p015}
If $\{v_1,v_2\}, \{v_2,v_3\}, \{v_1,v_4\} \in \mathbb{D}_u$ and
$\{v_1,v_2\}$ dominates $v_3$, then $\{v_1,v_4\}$ dominates $v_3$.
\end{lemma}
{\bf Proof:} According to the Lemma~\ref{ddom_p01}, either $\{v_1,v_4\}$
dominates $v_2$, or $\{v_1,v_2\}$ dominates $v_4$. This implies that one of the
two following cases are possible:
\begin{itemize}
\item[(1)] $\{v_1,v_4\}$ dominates $\{v_1,v_2\}$,
\item[(2)] $\{v_1,v_2\}$ dominates $\{v_1,v_4\}$.
\end{itemize}

\noindent {\bf Case 1:}
If $\{v_1,v_4\}$ dominates $\{v_1,v_2\}$, then from the condition of the Lemma by transitivity of
dominator relation it follows that $\{v_1,v_4\}$ dominates $v_3$.

\noindent {\bf Case 2:}
If $\{v_1,v_2\}$ dominates $\{v_1,v_4\}$, then by the antisymmetry of
dominator relation it follows that $\{v_1,v_4\}$ does not dominate
$\{v_1,v_2\}$. The vertex $v_1$ is dominated by $\{v_1,v_4\}$, thus
$v_2$ is not dominated by $\{v_1,v_4\}$.

Since $\{v_1,v_2\}$ dominates $v_3$, it implies that $\{v_1,v_2\}$ dominates
$\{v_2,v_3\}$, thus $\{v_2,v_3\}$ does not dominate $\{v_1,v_2\}$. The
vertex $v_2$ is dominated by $\{v_2,v_3\}$, thus $v_1$ is not
dominated by $\{v_2,v_3\}$.

Since $v_2$ is not dominated by $\{v_1,v_4\}$ and $v_1$ is not
dominated by $\{v_2,v_3\}$, according to the Lemma~\ref{ddom_p04} $\{v_1,v_4\}$
dominates $v_3$.
\begin{flushright}
$\Box$
\end{flushright}


\begin{lemma} \label{ddom_p05}
For all $\{v_1,v_2\} \in \mathbb{D}_u$ and for all $\{v_3,v_4\} \in
\mathbb{D}_u$, there exist $\{v_5,v_6\} \in \mathbb{D}_u$ such that $\{v_1,v_2\}$
dominates $\{v_5,v_6\}$ and $\{v_3,v_4\}$ dominates $\{v_5,v_6\}$.
\end{lemma}
{\bf Proof:}
Three cases are possible:
\begin{itemize}
\item[(1)] $\{v_1,v_2\}$ and $\{v_3,v_4\}$ have two common
vertices, i.e they are the same set.
\item[(2)] $\{v_1,v_2\}$ and $\{v_3,v_4\}$ have one common
vertex,
\item[(3)] $\{v_1,v_2\}$ and $\{v_3,v_4\}$ do not have common
vertices.
\end{itemize}

We prove the Lemma by identifying the dominator set $\{v_5,v_6\}$
for all three cases.

\noindent {\bf Case 1:}
The Lemma trivially holds by choosing $\{v_1,v_2\}$ to be $\{v_5,v_6\}$.

\noindent {\bf Case 2:}
Suppose that $v_2$ is the common vertex, i.e. the second immediate
dominator is $\{v_2,v_3\}$. According to the Lemma~\ref{ddom_p01}, $\{v_1,v_2\}$
dominates $v_3$ or $\{v_2,v_3\}$ dominates $v_1$. Without any loss of
generality, assume that $\{v_1,v_2\}$ dominates $v_3$. It immediately
follows that $\{v_1,v_2\}$ dominates $\{v_2,v_3\}$. Thus the Theorem
holds by choosing $\{v_2,v_3\}$ to be $\{v_5,v_6\}$.

\noindent {\bf Case 3:}
If one dominator dominates the other one, then the Theorem holds by
choosing the dominated dominator to be $\{v_5,v_6\}$.

Assume that none of the dominators dominates each other. It means at
least one vertex in both dominators is not dominated by the other
dominator.  Note that with current assumption it is impossible that
two vertices in any of the dominators are not dominated by the other
dominator, since it would contradict Lemma~\ref{ddom_p04}. Thus exactly
one vertex from both dominators is not dominated by the other dominator
and no other cases are possible.

Without any loss of generality, assume that $\{v_3,v_4\}$ does not dominate
$v_1$ and $\{v_1,v_2\}$ does not dominate $v_4$. According to
the Lemma~\ref{ddom_p03}, $\{v_2,v_3\}\in \mathbb{D}_u$. According to the Lemma~\ref{ddom_p04},
$\{v_3,v_4\}$ dominates $v_2$, thus $\{v_3,v_4\}$ dominates
$\{v_2,v_3\}$. Also $\{v_1,v_2\}$ dominates $v_3$, thus $\{v_1,v_2\}$
dominates $\{v_2,v_3\}$. The Lemma holds by choosing $\{v_2,v_3\}$ to be
$\{v_5,v_6\}$.
\begin{flushright}
$\Box$ \\
\end{flushright}

\begin{lemma} \label{ddom_p06}
For any non-empty subset $\mathbb{A}$ of $\mathbb{D}_u$, there exist $\{v_1,v_2\} \in
\mathbb{D}_u$ such that $\{v_1,v_2\}$ dominated by all dominators in $\mathbb{A}$.
\end{lemma}

\noindent {\bf Proof:}
We prove the Lemma by induction on the size of the set $\mathbb{A}$.

\noindent {\bf Basis:}
If $|\mathbb{A}|=1$, then the dominator which is dominated by all dominators in
$\mathbb{A}$ is the dominator which constitutes $\mathbb{A}$, i.e. $\{v_1,v_2\} \in A$.

\noindent {\bf Inductive step:}
Assume the Lemma holds for $|\mathbb{A}|=k$. Next we show that the Lemma holds
for $|\mathbb{A}|=k+1$, where $k \in \{ 1,2, \ldots, ,|\mathbb{D}_u-1|\}$.

Let $\mathbb{B}$ be a proper subset of $\mathbb{A}$ such that $|B|=k$. Since $\mathbb{A}$ is
a subset of $\mathbb{D}_u$, $\mathbb{B}$ is a subset of $\mathbb{D}_u$ as well. According to
the assumption, there exists $\{v_3,v_4\} \in \mathbb{D}_u$ such that $\{v_3,v_4\}$
is dominated by all vertices in $\mathbb{B}$.  Let $\{v_5,v_6\}$ be the remaining
dominator of $\mathbb{A}$ which does not belong to $\mathbb{B}$, i.e. $\{v_5,v_6\} \in
A-B$.  According to the Lemma~\ref{ddom_p05}, there exists $\{v_7,v_8\} \in \mathbb{D}_u$
such that $\{v_3,v_4\}$ dominates $\{v_7,v_8\}$ and $\{v_5,v_6\}$
dominates $\{v_7,v_8\}$.  All dominators in $\mathbb{B}$ dominate $\{v_3,v_4\}$
and $\{v_3,v_4\}$ dominate $\{v_7,v_8\}$, thus, using transitivity of
dominator relation, all dominators in $\mathbb{B}$ dominate
$\{v_7,v_8\}$. Since $\{v_5,v_6\}$ dominates $\{v_7,v_8\}$ as well, we
can conclude that all dominators in $\mathbb{A}$ dominate $\{v_7,v_8\}$. Thus,
$\{v_1,v_2\}=\{v_7,v_8\}$.
\begin{flushright}
$\Box$ \\
\end{flushright}

Finally, we prove that immediate double-vertex dominators are unique.
As we have shown in Section~\ref{ddom_prelim}, this property does not extend
to the dominators of a larger size.

\begin{theorem} \label{ddom_p08}
For any $u  \in \mathbb{V}$, if $\mathbb{D}_u$ is non-empty, then there exist a unique immediate double-vertex
dominator of $u$.
\end{theorem}

\noindent {\bf Proof:} It immediately follows from the Lemma~\ref{ddom_p06} that there exists
$\{v_1,v_2\} \in \mathbb{D}_u$ such that $\{v_1,v_2\}$ is dominated by all
dominators in $\mathbb{D}_u$. Due to the antisymmetry of dominator
relation, $\{v_1,v_2\}$ does not dominate any other dominator in
$\mathbb{D}_u$. By Definition~\ref{ddom_idom}, $\{v_1,v_2\}$ is an immediate
double-vertex dominator of $u$.

To prove the uniqueness of the immediate double-vertex dominator, assume there is another immediate
double-vertex dominator $\{v_3,v_4\} \in \mathbb{D}_u$. Since any dominator
in $\mathbb{D}_u$ dominates $\{v_1,v_2\}$, it means that $\{v_3,v_4\}$ dominates
$\{v_1,v_2\}$. This contradicts the Definition~\ref{ddom_idom}.
\begin{flushright}
$\Box$ \\
\end{flushright}


\section{A Data Structure for Representing Dominators}
\label{ddom_ds}

In this section, we describe a data structure for representing all double-vertex dominators of a given 
vertex in linear space\footnote{A preliminary short version of the paper presenting this data structure appeared in the Proceedings of the Design and Test in Europe Conference (DATE’2005)~\cite{TeD05b}.}.

Given one vertex in a double-vertex dominator $\{v,w\}$,  say $v$, we call
the other vertex $w$ a {\em matching} vertex of $v$ with respect to $u$.
A vertex may have more than one matching vertices with respect to $u$. 
We represent the set of all matching vertices of a vertex by the
following vector.

\begin{figure}[t!]
\begin{center}
\includegraphics*[width=4in]{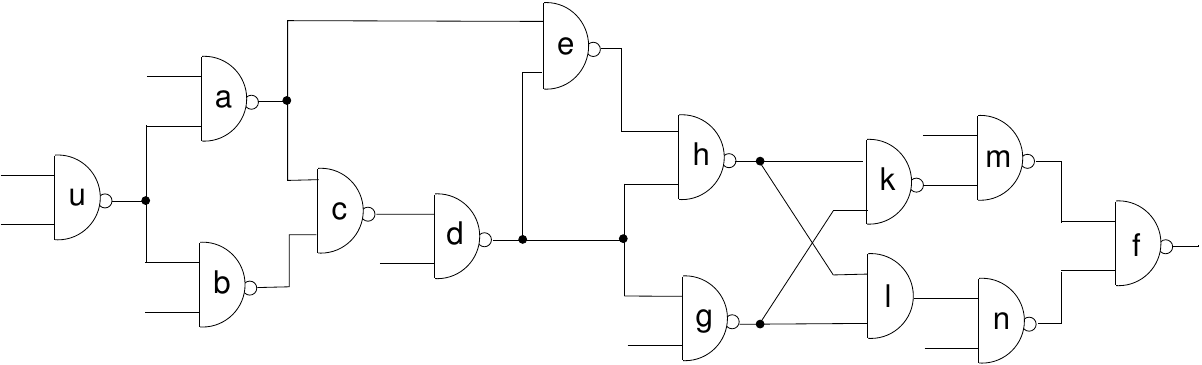}
\caption{An example circuit.} \label{ddom_f3}
\end{center}
\end{figure}

\begin{defn} \label{ddom_mv}
For any $v \in \mathbb{D}_u$, the {\em matching vector} of $v$ with respect to $u$, denoted by $M_u(v)$, consists of
all vertices $w \in \mathbb{V}$ such that $\{v,w\}$ is a double-vertex dominator of $u$.
The order of vertices in $M_u(v)$ is defined as follows: 
If $\{v,w\} \in \mathbb{D}_u$ dominates $\{v,w'\} \in \mathbb{D}_u$, then $w'$ precedes $w$ in $M_u(v)$.
\end{defn}

\begin{lemma} \label{ddom_tmv1}
For every $v \in \mathbb{D}_u$, there exist  a unique matching vector $M_u(v)$.
\end{lemma}
\noindent {\bf Proof:} The set of vertices which constitute $M_u(v)$ for a given $v \in \mathbb{D}_u$ is uniquely
determined by the Definition~\ref{ddom_mv}. It remains to prove that the
order of elements in $M_u(v)$ is unique. 

By Definition~\ref{ddom_mv}, the vertices of $\mathbb{D}_u$ are 
ordered according to the dominator relation. 
Given any pair of double-vertex dominators of $u$, say $\{v,w\}$ and  $\{v,w'\}$, 
by Lemma~\ref{ddom_p01}, either $\{v,w\}$ dominates $w'$,
or $\{v,w'\}$ dominates $w$. This implies that either $\{v,w\}$
dominates $\{v,w'\}$, or $\{v,w'\}$ dominates $\{v,w\}$. Thus, 
the order imposed by the dominator relation on the elements 
of $\mathbb{D}_u$ is total.
\begin{flushright}
$\Box$ \\
\end{flushright}

As an example, consider the circuit in Figure~\ref{ddom_f3}.
The set of all double-vertex dominators of 
$u$ is: $\mathbb{D}_u = \{\{a,b\}$, $\{a,c\}$, $\{a,d\}$, $\{e,c\}$, $\{e,d\}$,
$\{h,c\}$, $\{h,d\}$, $\{h,g\}$, $\{k,l\}$, $\{m,l\}$, $\{k,n\}$,
$\{m,n\}\}$. 
Therefore, we have the following matching vectors with respect to $u$:
\[
\begin{array}{l}
M_u(a) = (b,c,d) \\
M_u(b) = (a) \\
M_u(c) = (a,e,h) \\
M_u(d) = (a,e,h) \\
M_u(e) = (c,d) \\
M_u(g) = (h) \\
M_u(h) = (c,d,g) \\
M_u(k) = (l,n) \\
M_u(l) = (k,m) \\
M_u(m) = (l,n) \\
M_u(n) = (k,m) 
\end{array}
\]

Let $\mathbb{M}_u$ be the set of all matching vectors of all vertices in $\mathbb{D}_u$.
The set $\mathbb{M}_u$ can be partitioned into a set of connected components
which we call {\em clusters}.

\begin{defn} \label{ddom_deps}
A set of matching vectors $\mathbb{M}'_u \subseteq \mathbb{M}_u$ is a {\em cluster} if:
\begin{enumerate}
\item[(1)]
$\forall M_u(v) \in \mathbb{M}'_u$ and $\forall M_u(w) \in \mathbb{M}_u - \mathbb{M}'_u$, $M_u(v) \cap M_u(w) = \emptyset$
\item[(2)]
$\mathbb{M}'_u$ cannot be partitioned into two clusters satisfying (1).
\end{enumerate}
\end{defn}

In the example above, $\mathbb{M}_u$ can be partitioned into 4 clusters:
$\{M_u(a), M_u(e), M_u(h)\}$, $\{M_u(b), M_u(c), M_u(d), M_u(g)\}$,
$\{M_u(k), M_u(m)\}$, and $\{M_u(l), M_u(n)\}$.

Finally, we introduce a structure  
which will allow us to represent all clusters of $\mathbb{M}_u$ in linear space.

\begin{defn} \label{ddom_comp}
A vector $C(\mathbb{M}'_u)$  is the {\em composition vector} for a set of matching vectors $\mathbb{M}'_u \subseteq \mathbb{M}_u$, if: 
\begin{enumerate}
\item It contains each matching vector of $\mathbb{M}'_u$ as a subvector,
\item It contains only matching vectors from $\mathbb{M}'_u$,
\item It contains no duplicated vertices.  
\end{enumerate}
\end{defn}

\begin{theorem} \label{ddom_tmv2}
For any two vertices $v, v' \in \mathbb{D}_u$, it holds that either
\begin{enumerate}
\item $M_u(v) \cap M_u(v') =$ {\em suffix}$(M_u(v)) =$ {\em prefix}$(M_u(v'))$, or
\item $M_u(v) \cap M_u(v') =$ {\em suffix}$(M_u(v')) =$ {\em prefix}$(M_u(v))$.
\end{enumerate}
\end{theorem}
\noindent {\bf Proof:}  See Appendix A.

\begin{figure}[t!]
\begin{center}
\includegraphics*[width=2.8in]{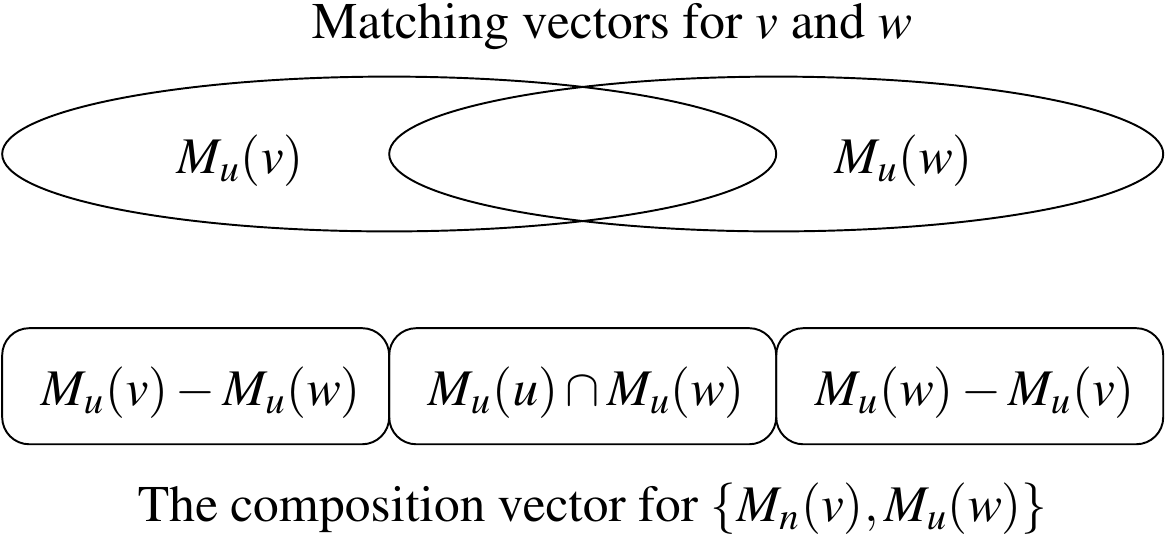}
\caption{The relation between two overlapping matching vectors and their composition vector.} \label{comp_vector}
\end{center}
\end{figure}

An obvious implication of the Theorem~\ref{ddom_tmv2} is that, for any two matching vectors, there exists a composition vector. Furthermore if the two matching vectors
have vertices in common, then the composition vector is unique (see Figure~\ref{comp_vector} for an illustration). It can also be shown that, 
for any set of matching vectors,  there exists a composition vector.

In the example above, $C(M_u(a), M_u(e), M_u(h)) = (b,c,d,g)$, $C(M_u(b), M_u(c), M_u(d),$ $M_u(g)) = (a,e,h), C(M_u(k), M_u(m)) =$ $(l,n)$, and $C(M_u(l), M_u(n)) = (k,m)$.
Note that the set of matching vectors of vertices of the first composition vector 
is equivalent to the second cluster, and vice verse. Similarly, the set of matching vectors of vertices of the third composition vector is equivalent to the fourth cluster, and vice verse. We call such clusters  {\em  complimentary}.

\begin{defn} \label{cluster_comp}
The cluster is {\em complimentary} to a cluster $\mathbb{M}'_u \subseteq \mathbb{M}_u$,
denoted by $\mathbb{\overline{M}}'_u$,
if the set of all matching vectors $M_u(v)$ of all $v \in C(\mathbb{M}'_u)$ constitute a cluster equivalent to $\mathbb{\overline{M}}'_u$.
\end{defn}

It is easy to show that if $\mathbb{\overline{M}}'_u$ is complimentary to $\mathbb{M}'_u$, then $\mathbb{M}'_u$ is complimentary to $\mathbb{\overline{M}}'_u$ as well. Each double-vertex dominator in $\mathbb{D}_u$ has one of its vertices in some cluster $\mathbb{M}'_u \subseteq \mathbb{M}_u$ and another vertex in $\mathbb{\overline{M}}'_u$. The following Lemma follows directly.

\begin{lemma}
The set $\mathbb{M}_u$ can be partitioned into pairs of complimentary clusters.
\end{lemma}

This brings us to the data structure for representing $\mathbb{D}_u$.

\begin{defn} \label{ddom_d1}
The set $\mathbb{D}_u$ of all double-vertex dominators of any $u \in \mathbb{V}$
can be represented by the {\em dominator chain} $\mathcal{D}(u)$ 
which is a vector of pairs of composition vectors of complimentary clusters of $\mathbb{M}_u$:
\[
\mathcal{D}(u) = (\{C(\mathbb{M}^1_u), C(\mathbb{\overline{M}}^1_u)\},  \ldots, \{C(\mathbb{M}^k_u), C(\mathbb{\overline{M}}^k_u)\}),
\]
where $\mathbb{M}^i_u$ is the $i$th cluster of $\mathbb{M}_u$, for $i \in \{1,2,\ldots,k\}$.
The order of clusters in $\mathcal{D}(u) $ is defined as follows: 
If $v$ is the first vertex of $C(\mathbb{M}^i_u)$ and $w$ is the first vertex of $C(\mathbb{\overline{M}}^i_u)$,
then $\{v,w\}$ dominates every vertex in $C(\mathbb{M}^j_u)$ and $C(\mathbb{\overline{M}}^j_u)$ for all $i < j \leq k$.
Each vertex $v$ which is contained in $\mathcal{D}(u) $ is associated
with a pair $(v_{min}, v_{max})$ representing of the first
and the last vertex of $M_u(v)$.
\end{defn}

For the circuit in Figure~\ref{ddom_f3}, the dominator chain for $u$ is
\[
\mathcal{D}(u) = (\{(a,e,h), (b,c,d,g)\}, \{(k,m), (l,n)\}),
\]
$a$ is associated with $(b,d)$, $b$ is associated with $(a,a)$, etc.

\section{Operations of Dominator Chains}

One of the tasks for which dominator chains are used in this paper is to
identify whether a given pair of vertices $\{v,w\}$ is a
double-vertex dominator of some vertex $u \in \mathbb{V}$ or not. 
Assume that we have the dominator chain $\mathcal{D}(u)$ and 
that pairs $(v_{min}, v_{max})$ consisting of the first
and the last vertex of $M_u(v)$ are associated with each $v \in \mathbb{D}_u$.
For each vertex $v' \not\in \mathbb{D}_u$,  we set $(v_{min}, v_{max}) = \emptyset$. 
Then, to determine whether $\{v,w\}$ is a
double-vertex dominator of $u$,  we first check whether $(v_{min}, v_{max})$ and $(w_{min}, w_{max})$ are empty.
If they are, $\{v,w\}$ is not a dominator of $u$. Otherwise,
we take $v_{min}$ and search for this vertex in $\mathcal{D}(u)$. The position of $v_{min}$ in 
$\mathcal{D}(u)$ gives us the starting point of $M_u(v)$. We need to traverse $M_u(v)$ until its last vertex, $v_{max}$,
to determine whether $w \in M_u(v)$.  If $w \in M_u(v)$, then $\{v,w\}$ is a dominator of $u$. Otherwise, $\{v,w\}$ is not a dominator of $u$. 
Such a procedure has a linear time complexity with respect to the size of $\mathcal{D}(u)$. 
However, ii can be further improved by indexing vertices of $\mathcal{D}(u)$ as follows.


We partition $\mathcal{D}(u)$ into two vectors $\mathcal{L}(u)$ ("left") and $\mathcal{R}(u)$ ("right").
For each pair of composition vectors $\{C(\mathbb{M}^i_u), C(\mathbb{\overline{M}}^i_u)\}$ in $\mathcal{D}(u)$,
we put all vertices of one composition vector in $\mathcal{L}(u)$  and all vertices of another composition vector  in $\mathcal{R}(u)$.
It does not matter whether we put all $C(\mathbb{M}^i_u)$ in $\mathcal{L}(u)$ and all $C(\mathbb{\overline{M}}^i_u)$ in 
$\mathcal{R}(u)$, or vice verse. However, once we make a choice for the first pair of  composition vectors in  $\mathcal{D}(u)$,
this choice should be followed for all  pairs in $\mathcal{D}(u)$. It is also possible to make $\mathcal{L}(u)$ and $\mathcal{R}(u)$
unique by imposing the topological order on vertices of the circuit graph. In this case,
we put $C(\mathbb{M}^i_u)$ in $\mathcal{L}(u)$  
if the first vertex of $C(\mathbb{M}^i_u)$ precedes the first vertex of $C(\mathbb{\overline{M}}^i_u)$.
Otherwise, we put $C(\mathbb{M}^i_u)$ in $\mathcal{R}(u)$.  

For the circuit in Figure~\ref{ddom_f3}, the dominator chain can be partitioned as follows:
\[
\begin{array}{l}
\mathcal{L}(u) = (a,e,h,k,m), \\
\mathcal{R}(u) = (b,c,d,g,l,n).
\end{array}
\]

To make possible a constant time look-up for dominators, three
parameters are assigned to vertices:
\begin{itemize}
\item
For all $v \in \mathbb{D}_u$ we assign $flag(v) \in \{$ {\em left,right} $\}$, which distinguishes
whether $v$ belongs to $\mathcal{L}(u)$ or $\mathcal{R}(u)$.  
\item
For all $v \in \mathcal{L}(u) (\mathcal{R}(u))$,  we assign $index(v)$ which
indicates the position of $v$ in $\mathcal{L}(u) (\mathcal{R}(u))$.  
\item
Instead of associating with
each $v \in \mathbb{D}_u$ a pair of vertices $(v_{min},v_{max})$, we associating with
each $v$  a pair of indexes $(min,max)$, where
$min(v)=index(v_{min})$, $max(v) = index(v_{max})$. 
\end{itemize}

In the example above, $flag(a) =$ {\em left}, $flag(b) =$ {\em right}, 
$index(b) = 1$, $index(c) = 2$, 
$(min(b),max(b)) = (1,1)$, $(min(c),max(c)) = (1,3)$, etc.

Now we can check whether $\{v,w\}$ dominates $u$ as
follows:
\begin{enumerate}
\item
Check if $flag(v) \not= flag(w)$.
If yes, go to step 2. Otherwise, $\{v,w\} \not\in \mathbb{D}_u$.
\item Check if $min(v) \leq index(w) \leq max(v)$. If yes,
$\{v,w\} \in \mathbb{D}_u$. Otherwise, $\{v,w\} \not\in \mathbb{D}_u$.
\end{enumerate}


\section{An Algorithm for Finding Dominators} \label{ddom_alg}

The algorithm presented in this section
takes as its input a circuit graph $G =(\mathbb{V},\mathbb{E},root)$ and a
vertex $u \in \mathbb{V}$.  It returns the dominator chain $\mathcal{D}(u)$.
The pseudo-code of the algorithm is shown in Figure~\ref{ddom_code_d}.

In order to construct $\mathcal{D}(u)$, the following steps are followed:
\begin{enumerate}
\item Find all single-vertex dominators of $u$.
\item Set $\mathcal{D}(u)=\emptyset$ and $v=u$.
\item Construct the dominator chain  $\mathcal{D}(v)$ for $v$ assuming that $ idom(v)$ is the sink
and append it to the end of $\mathcal{D}(u)$.
\item Set $v=idom(v)$ and repeat Step 3 until $v \neq root$.
\end{enumerate}

To simplify the description of the algorithm, we assume that there are
no single-vertex dominators of $u$ with respect to $root$, i.e.
we focus on the Steps 3 and 4. 



\begin{figure}[t!] 
\begin{center}
\parbox{0cm} 
{
\begin{tabbing}  
m \== m \== m \== m \== m \== m \== m \== \kill
\ALGORITHM $\procname{DominatorChain}(\mathbb{V},\mathbb{E},root,u)$ \\
\INPUT $\mathbb{V}$ is a set of vertices, $\mathbb{E} \subseteq \mathbb{V} \times \mathbb{V}, root \in \mathbb{V}, u \in \mathbb{V}$.\\
\> Construct a path $P_1 \subseteq V$ from $u$ to $root$;\\
\> Construct a path $P_2 \subseteq V$ from $u$ to $root$ such that $P_2 \cap P_1 = \{u,root\}$; \\
\> Construct a path $P_3 \subseteq V$ from $u$ to $root$ such that $P_3 \cap (P_1 \cup P_2) = \{u,root\}$; \\
\> \IF $P_3$ is constructed \THEN \\
\> \> \RETURN $\mathcal{D}(u)= \emptyset$; \\ 
\> \FOREACH $v \in \mathbb{V}$ \DO \\
\> \> Set {\em marked(v)} = 0;\\
\> \END \\
\> $\procname{AssignMinMax}(P_1,P_2)$;\\
\> \FOREACH $v \in \mathbb{V}$  \DO \\
\> \> Set {\em marked(v)} = 0;\\
\> \END \\
\> $\procname{AssignMinMax}(P_2,P_1)$;\\
\> $\mathcal{L}(u)=\procname{ConstructVector}(P_1,P_2)$;\\
\> $\mathcal{R}(u)=\procname{ConstructVector}(P_2,P_1)$;\\
\> $\procname{ConvertMinMax}(\mathcal{L}(u),P_2)$;\\
\> $\procname{ConvertMinMax}(\mathcal{R}(u),P_1)$;\\
\> \RETURN $\procname{ConstructD\_u}(\mathcal{L}(u),\mathcal{R}(u))$;\\
\END
\end{tabbing}
}
\caption{Pseudo-code of the presented algorithm for finding double-vertex dominators of a vertex $u$.}\label{ddom_code_d}
\end{center}
\end{figure}

The presented algorithm exploits the following property of disjoint paths.
Recall that we call two paths disjoint if the intersection of
sets of their non-terminal vertices is empty.

\begin{lemma} \label{ddom_al01}
If there are two disjoint paths from $u$ to
$root$, $P_1$ and $P_2$, then, for any double-vertex dominator $\{v,w\}$ of $u$, it
holds that $v \in P_1$ and $w \in P_2$.
\end{lemma}
{\bf Proof:} By Definition~\ref{ddom_dom2}, 
at least one vertex of the double-vertex dominator $\{v,w\}$
should be present in any path from $u$ to $root$.
Since $P_1$ and $P_2$ are disjoint,
none of their vertices belong to both paths
except $u$ and $root$. Vertices $u$ and $root$ are single vertex
dominators of $u$, thus they do not belong to $\mathbb{D}_u$. Therefore, one
vertex of the pair $\{v,w\}$ should belong to $P_1$ and another one to $P_2$.
\begin{flushright}
$\Box$ \\
\end{flushright}

It directly follows from the Lemma~\ref{ddom_al01} that 
if there exists a third path from $u$ to
$root$ which is disjoint with both $P_1$ and $P_2$, then $u$ has 
no double-vertex dominators. We use this property to
bound the search space for double-vertex dominators.

\begin{figure}[t!]
\begin{center}
\parbox{0cm} 
{
\begin{tabbing}
m \== m \== m \== m \== m \== m \== m \== \kill
\ALGORITHM $\procname{AssignMinMax}(P_1,P_2)$ \\
\INPUT $P_1 = (v_1=u , v_2, v_3, \ldots, v_{|P_1|}=root)$,\\
~~~~~~~~$P_2 = (w_1=u , w_2, w_3, \ldots, w_{|P_2|}=root)$.\\
\> $reached\_P_1 = 0$; \\
\> $reached\_P_2 = 1$; \\
\> $new\_reached\_P_1=reached\_P_1$; \\
\> $new\_reached\_P_2=reached\_P_2$; \\
\> $last\_prime=0$; \\
\> \FOREACH $i$ from 1 to $|P_1|-1$ \DO \\
\> \> \IF $reached\_P_1 > i$ \THEN \\
\> \> \> /*By setting $min(v_i) = |P_2|$ we remove $v_i$ from*/ \\
\> \> \> /*the list of potential candidates into dominators*/ \\
\> \> \> $min(v_i) = |P_2|$; \\
\> \> \> $prime(v_i)=last\_prime$;\\
\> \> \ELSE \\
\> \> \> $min(v_i) = reached\_P_2$; \\
\> \> \> $prime(v_{last\_prime})=i$;\\
\> \> \> $last\_prime=i$;\\
\> \> $\procname{FindReachable}(v_i,P_1,P_2)$; \\
\> \> \IF $reached\_P_1 < new\_reached\_P_1$ \THEN \\
\> \> ~~~~~~~~~~~~~~~~~~~~~~~~~~~~~ $reached\_P_1=new\_reached\_P_1$  \\
\> \> \IF $reached\_P_2 >= new\_reached\_P_2$ \THEN \BREAK  \\
\> \> \FOREACH $j$ from $reached\_P_2$ to $new\_reached\_P_2-1$ \DO \\
\> \> \> $max(w_ j) = i$; \\
\> \> \END \\
\> \> $reached\_P_2= new\_reached\_P_2$; \\
\> \END \\
\> $prime(v_{last\_prime})=|P_1|$;\\
\END
\end{tabbing}
}
\caption{Pseudo-code of the procedure $\procname{AssignMinMax}$.}\label{ddom_code2}
\end{center}
\end{figure}

We search for three disjoint paths from $u$ to
$root$ using a modified version of the max-flow algorithm which 
operates on vertex rather than edge capacities~\cite{TeD04c}. 
The max-flow algorithm attempts to construct three augmenting
paths with $u$ as the source and $root$ as the sink. 
Each vertex is assigned a unit capacity. 
The net flow through each vertex should be either one or zero.
Therefore, the resulting augmenting paths are mutually disjoint by
construction. 

If the algorithm succeeds to find three disjoint  paths,
then by Lemma~\ref{ddom_al01}, $\mathbb{D}_u=\emptyset$.
If only two disjoint  paths are found, then we conclude that vertices on
these paths are potential candidates for $\mathbb{D}_u$. The
Lemma below helps us to distinguish which of them 
can belong to $\mathbb{D}_u$ and which are not.


\begin{lemma} \label{ddom_al03}
Let $P_1 = (v_1=u , v_2, v_3, \ldots, v_{|P_1|=root})$ and $P_2$ be two disjoint paths from $u$ to $root$. 
If there exists a path $P_3$ which starts at some vertex $v_i \in P_1$,
ends at some vertex $v_j \in P_1$, $i,j \in \{1,2,\ldots,|P_1|\}$,
and has not other common vertices with
neither $P_1$ nor $P_2$, then 
$v_k \not\in \mathbb{D}_u$ for all $v_k \in P_1$ such that $i < k < j$.
\end{lemma}
{\bf Proof:} 
The path $P_1$ can be seen as concatenation of three paths
$P_1 = P_4 P_5 P_6$ where $P_4$ is a prefix of $P_1$ having $v_i$
as its last vertex, $P_6$ is a suffix of $P_1$ having $v_j$
as its first vertex, and $P_5$ is the middle part of $P_1$ containing
all vertices from $v_i$ to $v_j$.
Denote by $P_7$ a path $P_7 = P_4 P_3 P_6$.

Consider some vertex $v_k \in P_5$. 
Since $v_k \in P_5$ and $v_k$ cannot appear twice in 
$P_1$, $v_k \not \in P_4$ and $v_k \not \in P_6$. Since
$P_1$ and $P_3$ have no common vertices except $v_i$ and $v_j$,
we can conclude that $v_k \not \in P_3$. This implies that $v_k \not \in P_7$, and
also that $v_k \not \in P_2$, because $P_1$ and 
$P_2$ are disjoint. 
Since paths $P_2$ and $P_7$ are two disjoint paths from $u$ to
$root$ and $v_k$ does not belong to any of them, by Lemma~\ref{ddom_al01},
$v_k \not \in \mathbb{D}_u$.
\begin{flushright}
$\Box$ \\
\end{flushright}


We call a vertex $v \in \mathbb{V}$ {\em prime} if any path from 
an ancestor of $v$ to a descendant of $v$ contains $v$.
By the Lemma~\ref{ddom_al03}, 
any pair of prime vertices $\{v,w\}$ such that $v \in P_1$ and $w \in P_2$, 
and $P_1$ and $P_2$ are disjoint, can potentially be a double-vertex dominator of $u$. 
The next Lemma put additional restrictions of pairs of vertices that can belong to
$\mathbb{D}_u$.

\begin{lemma} \label{ddom_al04}
Let $P_1 = (v_1=u , v_2, v_3, \ldots, v_{|P_1|}=root)$ and $P_2 = (w_1=u , w_2, w_3, \ldots, w_{|P_2|}=root)$ 
be two disjoint paths from $u$ to $root$. 
If there exists a path $P_3$ which starts at some vertex $v_i \in P_1$,
ends at some vertex $w_j \in P_2$, and has not other common vertices with
$P_1$ and $P_2$, then all pairs of vertices $\{v_k,w_l\}$ such that $i < k \leq |P_1|$ 
and $1 \geq l < j$ are not in $\mathbb{D}_u$.
\end{lemma}
{\bf Proof:} The path $P_1$ can be seen as a concatenation of two paths
$P_1=P_4P_5$ where $P_4 = (v_1,\ldots,v_i)$ and $P_6 = (v_i,\ldots,v_{|P_1|})$.  
Similarly, the path $P_2$ can be seen as a concatenation of two
paths $P_2=P_6P_7$ where $P_6 = (w_1,\ldots,w_j)$ and $P_7 = (w_i,\ldots,w_{|P_2|})$.
Denote by $P_8$ a path $P_8=P_4P_3P_7$.

Since $P_1$ and $P_3$ have no common vertices except $v_i$, 
we can conclude that, for any $i < k \leq |P_1|$, $v_k \not\in P_3$. Similarly, 
for any $1 \geq l < j$, $w_l\ \not\in P_3$ 
because $P_2$ and $P_3$ have no common vertices
except $w_j$.

Since, for any $i < k \leq |P_1|$, $v_k \not\in P_3$ and $v_k$ cannot appear twice in  
$P_1$, $v_k \not \in P_4$. Also, for any $1 \geq l < j$, $w_l\ \not \in P_4$ because
$P_1$ is disjoint with $P_2$.

Similarly,  since for any $1 \geq l < j$, $w_l\ \not \in P_6$  
and $w_l$ cannot appear twice in  
$P_2$, $w_l\ \not \in P_7$. Also, for any $i < k \leq |P_1|$, $v_k\not \in P_7$ because
$P_1$ is disjoint with $P_2$.

It follows from above that, for any $i < k \leq |P_1|$ 
and $1 \geq l < j$, $v_k,w_l \not \in P_8$. Since $P_8$
is a path from $u$ to $root$, by the Definition~\ref{ddom_dom2}
that $\{v_k,w_l\} \not \in \mathbb{D}_u$.
\begin{flushright}
$\Box$ \\
\end{flushright}

\begin{figure}[t!]
\begin{center}
\parbox{0cm} 
{
\begin{tabbing}
m \== m \== m \== m \== m \== m \== m \== \kill
\ALGORITHM $\procname{FindReachable}(x,P_1,P_2)$ \\
\INPUT $x \in \mathbb{V}$, $P_1 = (v_1=u , v_2, v_3, \ldots, v_{|P_1|}=root)$,\\
~~~~~~~~$P_2 = (w_1=u , w_2, w_3, \ldots, w_{|P_2|}=root)$.\\

\> Let $P_1 = (v_1=u , v_2, v_3, \ldots, v_{|P_1|}=root)$;\\
\> Let $P_2 = (w_1=u , w_2, w_3, \ldots, w_{|P_2|}=root)$;\\
\> \FOREACH $y \in$ {\em TransFanout}$(x)$ \DO  \\
\> \> \IF $marked(y) = 1$ \THEN \BREAK \\
\> \> $marked(y) = 1$;\\
\> \> \IF $y = root$ \THEN \RETURN $(|P_1|,|P_2|)$; \\
\> \> \IF $y = v_{i}$ \THEN \\
\> \> \> \IF $i > new\_reached\_P_1$ \THEN $new\_reached\_P_1 = i$; \\
\> \> \> \BREAK \\
\> \> \IF $(y = w_{j})$ \THEN \\
\> \> \> \IF $n > new\_reached\_P_2$ \THEN $new\_reached\_P_2 = j$; \\
\> \> \> \BREAK \\
\> \END \\
\END
\end{tabbing}
}
\caption{Pseudo-code of the  procedure $\procname{FindReachable}$.}\label{ddom_code3}
\end{center}
\end{figure}

We use fields $max$ and $min$ of vertices of $P_1$ and $P_2$
to keep track of potential double-vertex dominators during the execution of the algorithm\footnote{Note that, because we
re-use the fields $max$ and $min$, their intermediate values during the execution of the algorithm might not be
in accordance with the definition in Section~\ref{ddom_ds}. The final values of $max$ and $min$ fields are set by the procedure $\procname{ConvertMinMax}$ before the termination of  the algorithm.}. 
If the field $max(v_k)$ of some $v_k \in P_1$  is assigned to $max(v_k) = i$, that means that we have identified that $\{v_k,w_j\} \not\in \mathbb{D}_u$ for all $w_j \in P_2$ such that $j > i$.
Similarly, if the field $max(v_k)$ of $v_k \in P_1$ is assigned to $max(v_k) = i$, then we have identified that 
$\{v_k,w_j\} \not\in \mathbb{D}_u$ for all $w_j \in P_2$ such that $j < i$.

The rules for assigning $max$ and $min$ fields follow from the Lemma~\ref{ddom_al04}. 
If there exist a  path $P_3 = (v_i, \ldots, w_j)$, $v_i \in P_1$, $w_j \in P_2$, disjoint with $P_1$ and $P_2$,
then $max(w_k) \leq i$ for all $k$ such that $1 < k \leq j-1$
and $min(v_l) \geq j$ for all $l$ such that $k+1 \geq l < |P_1|$.
Note that we write an inequality sign because there might be another path
$P_4 = (v_m, \ldots, w_n)$ disjoint with $P_1$ and $P_2$ such that $m < i$ and $n > j$.
In this case, $max(w_k) \leq m$ and $min(v_l) \geq n$. 
All paths disjoint with $P_1$ and $P_2$ should be considered to determine which indexes should be assigned to 
$max$ and $min$ fields. The following property summarizes the rules for assigning $max$ and $min$ fields.

\begin{prop} \label{ddom_minmax}
Let $P_1$ and $P_2$ be two disjoint paths from $u$ to $root$. 
Let $P_3 = (v_i, \ldots, w_j)$, $v_i \in P_1$, $w_j \in P_2$, be a path
disjoint with $P_1$ and $P_2$. Then:
\begin{itemize}
\item[(a)]
$max(v_k) = i$, $\forall w_k \in P_2$ such that $k < j$, 
where $i$ is the minimal index of a vertex of $P_1$ for which 
the path $P_3$ exists.
\item[(b)]
$max(v_k) = j$, $\forall v_k \in P_1$ such that $k > i$, 
where $j$ be the maximal index of a vertex of $P_2$ for which 
the path $P_3$ exists.
\end{itemize}
\end{prop}

\begin{figure}[t!]
\begin{center}
\parbox{0cm} 
{
\begin{tabbing}
m \== m \== m \== m \== m \== m \== m \== \kill
\ALGORITHM $\procname{ConstructVector}(P_1,P_2)$ \\
\INPUT $P_1 = (v_1=u , v_2, v_3, \ldots, v_{|P_1|}=root)$,\\
~~~~~~~~$P_2 = (w_1=u , w_2, w_3, \ldots, w_{|P_2|}=root)$.\\
\> $index\_count=1$;\\
\> $\mathcal{V}(u)=\emptyset$;\\
\> \FOREACH $i$ from 2  to $|P_1|-1$ \DO \\
\> \> $min=min(v_i)$;\\
\> \> $max=max(v_i)$;\\
\> \> \IF $min = |P_2|$ \THEN \BREAK \\
\> \> \IF $min(w_{min}) = |P_1|$ \THEN \\
\> \> \> /*min field is set to the index of the closest*/ \\
\> \> \> /*prime descendant of $w_{min}$ in $P_2$ */ \\
\> \> \>  $min(v_i)=prime(w_{prime(w_{min})})$;\\
\> \> \IF $min(w_{max}) = |P_1|$ \THEN \\
\> \> \> /*max field is set to the index of the closest*/ \\
\> \> \> /*prime ancestor of $w_{max}$ in $P_2$*/ \\
\> \> \>  $max(v_i)=prime(w_{max})$;\\
\> \> \IF $min(v_i) <= max(v_i)$ \THEN \\
\> \> \> Append $v_i$ to the end of vector $\mathcal{V}(u)$;\\
\> \> \> $index(v_i)=index\_count$;\\
\> \> \> $index\_count=index\_count+1$;\\
\> \END \\
\> \RETURN $\mathcal{V}(u)$;\\
\END
\end{tabbing}
}
\caption{Pseudo-code of the procedure $\procname{ConstructVector}$.}\label{ddom_code3a}
\end{center}
\end{figure}

\begin{figure}[t!]
\begin{center}
\parbox{0cm} 
{
\begin{tabbing}
m \== m \== m \== m \== m \== m \== m \== \kill
\ALGORITHM $\procname{ConvertMinMax}(\mathcal{V}(u),P_2)$ \\
\INPUT $\mathcal{V}(u) \subseteq V$, $P_2 = (w_1=u , w_2, w_3, \ldots, w_{|P_2|}=root)$.\\
\> \FOR all $v \in \mathcal{V}(u)$ \DO \\
\> \> $min(v) = index(w_{min(v)})$;\\
\> \> $max(v) = index(w_{max(v)})$;\\
\> \END \\
\END
\end{tabbing}
}
\caption{Pseudo-code of the procedure $\procname{ConvertMinMax}$.}\label{ddom_code4}
\end{center}
\end{figure}

The procedure $\procname{AssignMinMax}(P_1,P_2)$, shown in Figure~\ref{ddom_code2},  
allocates $max(v_i)$ field
for all vertices $v_i \in P_1$ and $min(w_j)$ field for all vertices $w_j
\in P_2$. This procedure also checks whether vertices of $P_1$ are
prime or not.  If $v_i  \in P_1$ is not a prime, then its field $prime(v_i)$ is set
to the index of the closest prime ancestor of $v_i$ in $P_1$.
If $v_i  \in P_1$ is a prime, then its field $prime(v_i)$ is set
to the index of the closest prime descendant of $v_i$ in $P_1$.

The main loop of the procedure $\procname{AssignMinMax}(P_1,P_2)$
iterates through all vertices $v_i$ of $P_1$ from the source to
the sink of $P_1$. For every $i$, in the beginning of the main loop, the
variable $reached\_P_1$ contains the maximum index of a vertex of
$P_1$ that can be reached from an ancestor of $v_i$ in $P_1$ by
a path disjoint with $P_1$ and $P_2$. Similarly, the variable $reached\_P_2$ contains the
maximum index of a vertex of $P_2$ that can be reached from
an ancestor of $v_i$ in $P_1$ by a path disjoint with $P_1$ and $P_2$.

In the main loop, first we check whether $v_i$ is prime or not. 
If $reached\_P_1 > i$, it means that there exists a
path $P_3$ from an ancestor of $v_i$ in $P_1$ to a 
descendant of $v_i$ in $P_1$ which is disjoint with $P_1$ and $P_2$. Thus by Lemma~\ref{ddom_al03} $v_i$ is not
prime. If $reached\_P_1 \leq i$ then no such path
exists and $v_i$ can be declared prime. According to the
Property~\ref{ddom_minmax}b $min(v_i)$ is set to $reached\_P_2$.

The procedure
$\procname{FindReachable}$, described later in this section, is used to update a pair
of global variables $new\_reached\_P_1$ and $new\_reached\_P_2$. The
values $new\_reached\_P_1$ represents the maximum index of a vertex of
$P_1$ that can be reached from $v_i$ or any ancestor of $v_i$ in $P_1$ 
by a path disjoint with $P_1$ and $P_2$. Since $v_i$ is an ancestor of
$v_{i+1}$, $new\_reached\_P_1$ represents the value of
$reached\_P_1$ for the next iteration of main loop. Similarly, the value
$new\_reached\_P_2$ represents the maximum index of a vertex of $P_2$
that can be reached from $v_i$ or any of its ancestors in $P_1$ by a
path with is disjoint with $P_1$ and $P_2$. Thus, $new\_reached\_P_2$ represents the value of
$reached\_P_2$ for the next iteration of the main loop.

If $new\_reached\_P_2 > reached\_P_2$, this means that, for every vertex $w_j \in P_2$ in
the range $(w_{reached\_P_2}, \ldots, w_{new\_reached\_P_2-1})$,
$i$ is the minimum index of a vertex in $P_1$ for which there exists a to a descendant of $w_j$ in $P_2$ which is disjoint with $P_1$ and $P_2$. According to the Property~\ref{ddom_minmax}a $max(w_j)$ is set to $i$.

The procedure $\procname{FindReachable}(x,P_1,P_2)$ sets
$marked(y)=1$ for all vertices $y$ which are reachable by
path which is disjoint with $P_1$ and $P_2$ from a given vertex $x$ and updates global variables
$new\_reached\_P_1$ and $new\_reached\_P_2$. The marking 
is performed by a depth-first search. Any
path disjoint with $P_1$ and $P_2$ which contains $y \not \in P_1 \cap P_2$ can be
extended to any of the vertices in the fanout of $y$. Such an extended path 
is disjoint with $P_1$ and $P_2$ as well.  So, all vertices in the fanout of $y$ are reachable by  paths disjoint with $P_1$ and $P_2$,
and therefore they are marked. $\procname{FindReachable}$ is called for
all newly marked vertices which do not belong to neither $P_1$ or $P_2$.

The maximum index of each marked vertex in a path $P_1$ ($P_2$) is stored in
the global variable $new\_reached\_P_1$ ($new\_reached\_P_2$). This variable
 represents the maximum index of a vertex of $P_1$ ($P_2$) that can be reached by 
a disjoint with $P_1$ and $P_2$ path from one of the vertices $x$ for which $\procname{FindReachable}(x,P_1,P_2)$ was
initially called.

The following theorem states that once all  fields $min$ and $max$ are set by
$\procname{AssignMin}$ $\procname{Max}(P_1,P_2)$ and
$\procname{AssignMinMax}(P_2,P_1)$, all remaining potential candidates to
double-vertex dominators are indeed double-vertex dominators.

\begin{theorem} \label{ddom_al02}
Let $P_1$ and $P_2$ be two disjoint paths from $u$ to $root$. 
If vertices $v_i \in P_1$ and $w_j \in P_2$ are prime, $max(w_j) \geq i$,
and $min(w_j) \leq i$, then $\{v_i,w_j\}$ is a double-vertex
dominator of $u$.
\end{theorem}
\noindent {\bf Proof:}  See Appendix B.

\begin{figure}[t!]
\begin{center}
\parbox{0cm} 
{
\begin{tabbing}
m \== m \== m \== m \== m \== m \== m \== \kill
\ALGORITHM $\procname{ConstructD\_u}(\mathcal{L}(u),\mathcal{R}(u))$ \\
\INPUT $\mathcal{L}(u) = (v_1, v_2, \ldots, v_{|\mathcal{L}(u)|})$, $\mathcal{L}(u) = (w_1, w_2, \ldots, w_{|\mathcal{R}(u)|})$. \\
\> $begin_L=1$; $begin_R=1$;\\
\> $end_L=1$; $end_R=1$;\\
\> $i=1$;\\
\> $\mathcal{D}(u) = \emptyset$;\\
\> \WHILE $end_L \neq |\mathcal{L}(u)|$ \DO \\
\> \> \WHILE 1 \DO\\
\> \> \> $end_{Rnew}=max(v_{end_L})$;\\
\> \> \> \IF $end_{Rnew} = end_R$ \THEN \BREAK \\
\> \> \> $end_R = end_{Rnew}$;\\
\> \> \> $end_{Lnew} = max(w_{end_R})$;\\
\> \> \> \IF $end_{Lnew} = end_L$ \THEN \BREAK \\
\> \> \> $end_L = end_{Lnew}$;\\
\> \> \END \\
\> \> Set $C(\mathbb{M}^i_u)=\{v_{begin_L}, \ldots, v_{end_L}\}$; /*$C(\mathbb{M}^i_u) \subseteq \mathcal{L}(u)$*/\\
\> \> Set $C(\mathbb{\overline{M}}^i_u)=\{w_{begin_R},\ldots, w_{end_R}\}$; /*$C(\mathbb{\overline{M}}^i_u)\subseteq \mathcal{R}(u)$*/\\
\> \> Append $\{C(\mathbb{M}^i_u), C(\mathbb{\overline{M}}^i_u)\}$ to $\mathcal{D}(u)$;\\
\> \> $i=i+1$;\\
\> \> $begin_L=end_L$;\\
\> \> $begin_R=end_R$;\\
\> \END \\
\end{tabbing}
}
\caption{Pseudo-code of the procedure $\procname{ConstructD\_u}$.}\label{ddom_code3b}
\end{center}
\end{figure}

The procedure $\procname{ConstructVector}(P_1,P_2)$ returns the vector
$\mathcal{V}(u)$, which is either $\mathcal{L}(u)$ or $\mathcal{R}(u)$. 
The vector $\mathcal{V}(u)$ consists
of a subset of vertices of $P_1$. According to the Theorem~\ref{ddom_al02},
a vertex $v_i$ belongs to $\mathcal{V}(u)$
 if there exists at least one prime vertex
in $P_2$ which is in the range between $min(v_i)$ and
$max(v_i)$. First, we check whether $min(v_i)$ and $max(v_i)$
contain indexes of prime vertices. If not, then they are updated as follows. The
field $min(v_i)$ is set to the minimum index of 
prime vertices $w_j$ in $P_2$ satisfying $j > min(v_i)$.
Similarly, the $max(v_i)$ is
set to the maximum index of 
prime vertices $w_j$ in $P_2$ satisfying $j < max(v_i)$.
Finally, if
$min(v_i) \leq max(v_i)$, then we can conclude that  $\{v_i,w_{min(v_i)}\}$ and
$\{v_i,w_{max(v_i)}\}$ are double-vertex
dominators of $u$ and  append $v_i$ at the end of $\mathcal{V}(u)$. 
At this point,  the position of $v_i$
in $\mathcal{V}(u)$ is known. Therefore, we set the index of $v_i$ to $index\_count$.
However, indexes of vertices $min(v_i)$ 
and $max(v_i)$ in the complimentary to $\mathcal{V}(u)$ 
vector of the dominator chain are not known yet. These indexes are assigned later by the
procedure $\procname{ConvertMinMax}(\mathcal{V}(u),P_2)$.

Finally, the dominator chain $\mathcal{D}(u)$ is constructed by the procedure
$\procname{ConstructD\_u}$ $(\mathcal{L}(u),\mathcal{R}(u))$. This procedure is optional, 
since for some applications it
is sufficient to find $\mathcal{L}(u)$ and $\mathcal{R}(u)$ along with $min(v)$,
$max(v)$ for all $v \in \mathbb{D}_u$.

The procedures $\procname{ConstructVector}(P_1,P_2)$,
$\procname{Convert}$ $\procname{MinMax}(\mathcal{V}(u),P_2)$ and
$\procname{ConstructD\_u}(\mathcal{L}(u),\mathcal{R}(u))$ have linear complexity with
respect to $|P_1|$, $|\mathcal{V}(u)|$, and $|\mathcal{L}(u)+\mathcal{R}(u)|$ respectively. The procedure
$\procname{FindReachable}(x,P_1,P_2)$ is called at most once for every
vertex during the call of
$\procname{AssignMinMax}(P_2,P_1)$. Each call of
$\procname{FindRe-}$ $\procname{achable}(x,P_1,P_2)$ iterates through all vertices in the fanout of $x$, thus $\procname{AssignMin}$ $\procname{Max}(P_2,P_1)$ has
linear time complexity with respect to the number of edges $\mathbb{E}$ in
the input graph.

Since all procedures of $\procname{DominatorChain}(\mathbb{V},\mathbb{E},root,u)$ 
have linear complexity
with respect to $|\mathbb{E}|$, the presented algorithm
has the complexity $O(|\mathbb{E}|)$. Its execution time is dominated by
the execution time of the procedures $\procname{AssignMinMax}(P_1,P_2)$ and
$\procname{AssignMinMax}(P_2,P_1)$. Therefore, the actual execution time of the presented algorithm is
proportional to $2|\mathbb{E}'|$, where $\mathbb{E}' \subseteq \mathbb{E}$ is the set
of edges in the transitive fanout of $u$.

\begin{table*}[t!]\centering
\begin{tabular}{|@{}c@{}|c|c|c||c|c|c||c|c|c|} \hline
            & 	      &     	                &  2-input       & All       & All      & Useful &\multicolumn{3}{c|}{Runtime, sec} \\ \cline{8-10}
Name  &  Inputs     & Outputs 	& AND gates  & 1-doms 	& 2-doms   & 2-doms  &\cite{DuTM04} & \cite{TeD05b} & presented  \\ \hline \hline

clma	&	94	&	115	&	24277	&	948	&	9819	&	2867	&	88.52	&	 0.41	 &	 0.34	\\
clmb	&	415	&	402	&	23906	&	361	&	8638	&	2356	&	98.09	&	 0.53	 &	 0.45	\\
mult32	&	64	&	96	&	10594	&	1150	&	27507	&	16442	&	885.62	 &	 2.98	 &	 1.45	\\
apex2	&	38	&	3	&	8755	&	853	&	1551	&	890	&	162.16	&	 0.23	 &	 0.16	 \\
too\_large	&	38	&	3	&	8746	&	971	&	2238	&	1467	&	136.02	 &	 0.22	 &	 0.14	\\
misex3	&	14	&	14	&	8155	&	59	&	2657	&	1224	&	29.83	&	 0.17	 &	 0.12	\\
seq	&	41	&	35	&	7462	&	1796	&	27631	&	13879	&	9.62	&	 0.25	 &	 0.16	\\
~cordic\_latches~	&	318	&	294	&	6212	&	7313	&	31714	&	12214	&	 4.27	 &	 0.36	&	0.28	 \\
bigkey	&	452	&	421	&	5661	&	2016	&	8822	&	2421	&	4.16	 &	 0.33	 &	 0.23	\\
s15850s	&	553	&	627	&	5389	&	27210	&	170189	&	31245	&	25.23	 &	 0.81	 &	 0.41	\\
alu4	&	14	&	8	&	5285	&	134	&	706	&	449	&	28.06	&	0.08	 &	 0.08	 \\
des	&	256	&	245	&	4733	&	3361	&	9231	&	2349	&	2.56	&	 0.25	 &	 0.17	\\
s15850	&	611	&	684	&	4172	&	34564	&	74941	&	16975	&	18.52	 &	 0.77	 &	 0.45	\\
apex5	&	114	&	88	&	3781	&	800	&	21728	&	8107	&	0.95	&	 0.17	 &	 0.12	\\
key	&	452	&	421	&	3537	&	1348	&	7717	&	2740	&	2.17	&	 0.28	 &	 0.19	\\
i8	&	133	&	81	&	3444	&	2068	&	8121	&	3296	&	0.83	&	 0.12	 &	 0.09	\\
ex1010	&	10	&	10	&	3278	&	0	&	545	&	92	&	11.33	&	0.14	 &	 0.14	 \\
dsip	&	452	&	421	&	2975	&	2245	&	6586	&	2059	&	1.75	 &	 0.23	 &	 0.2	\\
i10	&	257	&	224	&	2935	&	6446	&	81707	&	30608	&	4.95	&	 0.47	 &	 0.2	 \\
apex4	&	9	&	19	&	2905	&	0	&	841	&	165	&	8.7	&	0.12	&	 0.09	 \\
s13207s	&	483	&	574	&	2590	&	3179	&	13365	&	6673	&	2.28	 &	 0.22	 &	 0.16	\\
apex3	&	54	&	50	&	2419	&	1723	&	34386	&	29957	&	6.66	 &	 0.2	 &	 0.11	\\
C6288	&	32	&	32	&	2370	&	480	&	5743	&	3366	&	1.67	&	 0.27	 &	 0.2	 \\
C7552	&	207	&	108	&	2282	&	4604	&	87027	&	14728	&	19.12	 &	 0.31	 &	 0.11	\\
k2	&	45	&	45	&	2236	&	1827	&	16400	&	11693	&	5.42	&	 0.17	 &	 0.08	\\ \hline
{\bf total for 214}	&		&		&		& 177577	& 3777809	&	935309	&	1637.47	 &	 30.77  &  17.27 \\ \hline
\end{tabular}
\vspace*{1mm}
\caption{Benchmark results for IWLS'02 benchmark set.} \label{ddom_t1}
\end{table*}

\section{Experimental Results} \label{ddom_exp}

In this section, we compare the performance of the presented algorithm
to the algorithm for finding double-vertex dominators from~\cite{TeD05b}
and to the algorithm finding multiple-vertex dominators
from~\cite{DuTM04}.  
The algorithm~\cite{DuTM04} can compute
all $k$-vertex dominators of a given vertex for any $k$. In our experiment, 
we set $k$ to 2.

We have applied the three algorithms to 214 combinational benchmarks 
from the IWLS'02 benchmark set.
Table~\ref{ddom_t1} shows the results for 25 largest of these benchmarks. 
Columns 1, 2, 3 and 4 show the name of the benchmark, the number of
primary inputs, the number of primary outputs, and the number of 2-input AND gates in the benchmark,
respectively. 
In the last row of the Table~\ref{ddom_t1}, the $total$ is
computed for all 214 benchmarks.

In our experiments, we treated every primary output of a multiple-output
circuit as a separate function. Circuits for every primary output were extracted
from the original multiple-output circuit. For each resulting
single-output circuit, all dominators were computed for every primary
input with respect to the primary output. The numbers shown in Columns 5, 6
and 7 give the total number of dominators for all single
output circuits of the corresponding benchmark. 
The same dominator of several inputs was counted as one dominator.

In Column 5, we show the total number of single-vertex dominators
(except trivial dominators which are primary inputs and the primary output),
computed using the Lengauer and Tarjan's algorithm~\cite{LeT79}.

Column 6 shows the total number of double-vertex dominators computed by
 the presented algorithm, the algorithm ~\cite{TeD05b} and the algorithm~\cite{DuTM04}.  
All three algorithms found all double
vertex-dominators, therefore they produce the same result.
For most applications, useful dominators are
those which dominate more vertices then the size of the dominator
itself. Thus, in Column 7, we also show the number of all "useful"
double-vertex which dominate at least three primary inputs.

Columns 8, 9, and 10 show the runtime of three algorithms, in seconds. The
time was measured using the Unix command $time$ (user time). The
experiments were performed on a PC with a 1600 MHz AMD Turion64 CPU
and 1024 MByte main memory.

From Table~\ref{ddom_t1} we can see that the presented algorithm and
the algorithm~\cite{TeD05b} substantially outperform the algorithm~\cite{DuTM04}, delivering, on
average, an order of magnitude runtime reduction. This is not surprising since
they are specifically designed for double-vertex dominators.
We can also see that the presented algorithm consistently outperforms
the algorithm~\cite{TeD05b} on all 
benchmarks presented in Table~\ref{ddom_t1}.

In our implementation, the original benchmark circuits were converted to
an And-Inverter graph which consists of 2-input AND gates 
and Inverters~\cite{kuehlmann}. In such a graph, the majority of single vertex dominators have
the corresponding trivial double-vertex dominator (a pair of
vertices feeding the single-vertex dominator). The number of such
trivial double-vertex dominators can be roughly overapproximated to be
equal to the number of single-vertex dominators. Trivial double vertex
dominators  are usually less useful than the corresponding
single-vertex dominator. So, the numbers in Column 7 should be reduced by
the numbers in Column 5 to get a better picture of the number
of useful dominators.

Some rare circuits have less double-vertex dominators than
single-vertex dominators. Recall that our definition of multiple-vertex dominators
excludes redundancies. Therefore, in the extreme case of a tree-like
circuit with $n$ vertices the number of single-vertex dominators is $n$ while the number of double-vertex dominators is 0.

\section{Conclusion}
\label{ddom_con}
This paper presents supporting theory and algorithms for finding double-vertex dominators in directed acyclic graphs. Our results provide an efficient systematic way of partitioning a graph along the reconverging points of its signal flow. They might be useful in a number of CAD applications, including signal probability computation, switching activity estimation and cut
point identification. For example, in the method presented in~\cite{KhMKH01},
cut-points are used to progressively abstract a functional
representation by quantification.  Our dominator-based approach can
complement this method by providing a systematic way of identifying and selecting 
good cut-points for the abstraction.

Our results might also find potential applications beyond CAD borders.  In
general, any technique which use dominators in a directed acyclic graph 
might benefit from this work.

\bibliographystyle{IEEEtran}
\bibliography{dom}

\chapter*{Appendix}
{\bf A. Proof of the Theorem~\ref{ddom_tmv2}:}
Let $n$ denote the number of vertices common for $M_u(v)$ and $M_u(v')$. 
If $n=0$, then the Theorem~\ref{ddom_tmv2} holds trivially with vector
$M_u(v) \cap M_u(w)$ being empty.

Assume that $n>0$. We divide the prove into two parts. In the first
part, we prove that all $n$ common vertices should be in a suffix
of one vector, and in a prefix of the other one. In the second part, we
prove that the order of common vertices is the same in both vectors. 

\noindent {\bf Part 1:}
By assumption, there exists a common vertex, say $w  \in \mathbb{V}$, which
belong to both $M_u(v)$ and $M_u(v')$.
This implies that there exist dominators $\{v,w\} \in \mathbb{D}_u$ and
$\{v',w\} \in \mathbb{D}_u$. According to the Lemma~\ref{ddom_p01}, either
$\{v,w\}$ dominates $v'$ or $\{v',w\}$ dominates $v$.
This also means that either $\{v,w\}$ dominates $\{v',w\}$, or
$\{v',w\}$ dominates $\{v,w\}$. Without any loss of
generality, assume that $\{v,w\}$ dominates $\{v',w\}$.

First, we prove that a prefix of $M_u(v)$ whose last element is $w$
is always a subvector of $M_u(v')$ and a suffix of $M_u(v')$ whose first 
element is $w$ is always a subvector of $M_u(v)$.

Due to the antisymmetry of the dominator relation, $\{v',w\}$ does not
dominate $\{v,w\}$. Since $w$ is dominated by
$\{v',w\}$, thus $v$ is not dominated by $\{v',w\}$.

By the Definition~\ref{ddom_mv}, $\{v,w\}$ dominates
$\{v,w'\}$ for every vertex $w'$ preceding $w$ in $M_u(v)$.
Due to the antisymmetry of dominator relation, $\{v,w'\}$ does not dominate
$\{v,w\}$. Since $v$ is dominated by $\{v,w'\}$,
thus $w$ is not dominated by $\{v,w'\}$.

To summarize, we derived that there are dominators $\{v',w\} \in \mathbb{D}_u$ and
$\{v,w'\} \in \mathbb{D}_u$ such that $\{v',w\}$ does not dominate $v$ and
$\{v,w'\}$ does not dominate $w$. According to the
Lemma~\ref{ddom_p03}, this implies that $\{v',w'\} \in \mathbb{D}_u$. Therefore, 
every vertex $w'$ that precedes $w$ in $M_u(v)$
should also be contained in $M_u(v')$.

Using similar arguments as above, we can show that, for every vertex $w'$ succeeding $w$ in $M_u(v')$,
there exist dominators
$\{v,w\}$ and $\{v,w''\}$ such that $\{v,w\}$
does not dominate $w'$ and $\{v,w''\}$ does not dominate
$v$. Then, according to the Lemma~\ref{ddom_p03}, $\{v,w'\} \in \mathbb{D}_u$. 
This implies that every vertex $w'$ that succeeds $w$ in $M_u(v')$
should also be contained in $M_u(v)$.

By Lemma~\ref{ddom_p015}, the assumption that $\{v,w\}$ dominates $\{v',w\}$  implies that $\{v,w'\}$ dominates $\{v,w''\}$,
where $w'$ is any common vertex of $M_u(v)$ and $M_u(v')$. None of the common vertices can occupy a position $m$ in the vector
$M_u(v)$ such that $m > n$, since otherwise $m$ first vertices of $M_u(v)$ would be
contained in $M_u(v')$. This would contradict the fact that there are
only $n$ common vertices in both vectors. So, all $n$ common vertices
should be contained in a suffix of $M_u(v)$. Similarly, we can show
that all $n$ common vertices should be contained in a prefix of $M_u(v')$. 

\noindent {\bf Part 2:}
Next, we prove that $\{v,w\}$ dominating $\{v,w'\}$ implies that
$\{v',w\}$ dominates $\{v',w'\}$. This would imply the same
order of common vertices in vectors $M_u(v)$ and $M_u(v')$.

Assume that $\{v,w\}$ dominates $\{v,w'\}$.
Then using the same arguments as in the first part of the proof, we
can show that $\{w,v'\}$ does not
dominate $v$ and $\{w',v\}$ does not dominate $w'$. According to the
Lemma~\ref{ddom_p04}, $\{v',w\}$ dominates $w'$. This implies that
$\{v',w\}$ dominates $\{v',w'\}$.
\begin{flushright}
$\Box$ \\
\end{flushright}

{\bf B. Proof of the Theorem~\ref{ddom_al02}:} 
Assume that $\{v_i,w_j\}$ is not a double-vertex
dominator of $u$. Then there should be a path $P_3$ from $u$ to $root$ which
does not contain neither $v_i$ nor $w_j$. 

Define $P_4$ to be a vector containing all vertices of
$P_3$ which appear in either
$P_1$ or $P_2$. More formally, $x \in P_4$ if $x \in P_3$, and
either $x \in P_1$ or $x \in P_2$.  A vertex $x$ precedes a vertex $x'$ in
$P_4$ if $x$ precedes $x'$ in $P_3$.

Let $\mathbb{P}$ be a set containing all vertices that either precede $v_i$ in
$P_1$ or precede  $w_j$ in $P_2$.
Similarly, let $\mathbb{S}$ be a set of all vertices
that either succeed $v_i$ in $P_1$ or succeed $w_j$ in $P_2$.

Any vertex in $P_4$ belongs to either $\mathbb{P}$ or $\mathbb{S}$. Since
the first vertex of $P_4$, $u$, is in $\mathbb{P}$ and the last vertex of $P_4$,
$root$, is in $\mathbb{S}$, there exists $k$ such
that $x_k, x_{k+1} \in P_4$ and $x_k \in \mathbb{P}$ and $x_{k+1} \in \mathbb{S}$.
Let $P_5 = (x_k, \ldots, x_{k+1})$ be a subvector of $P_3$ containing all vertices of  $P_3$ from $x_k$ to $x_{k+1}$. 
By construction, $P_5$ does
not have any common vertices with neither $P_1$ nor $P_2$ except $u$ and $root$.

To summarize, from the assumption that $\{v_i,w_j\}$ is not a
double-vertex dominator $u$ we derived the existence of the path
$P_5$. Next we show that such a path $P_5$ cannot exist,
and therefore the assumption is not valid.

With respect to the source and the sink of $P_5$, there are four possible Cases:
\begin{enumerate}
\item $x_k \in P_1$ and $x_{k+1} \in P_1$,
\item $x_k \in P_2$ and $x_{k+1} \in P_2$,
\item $x_k \in P_1$ and $x_{k+1} \in P_2$,
\item $x_k \in P_2$ and $x_{k+1} \in P_1$.
\end{enumerate}

\noindent {\bf Case 1:}
If $P_5$ exists, then $v_i$ is not prime. This
contradicts the conditions of the Theorem~\ref{ddom_al02}.\\
{\bf Case 2:}
If $P_5$ exists, then $w_j$ is not prime. This contradicts
the conditions of the Theorem~\ref{ddom_al02}.\\
{\bf Case 3:}
If $P_5$ exists, then $max(w_j) \leq k$, where $k$ is the index of
$x_k$ in $P_1$. Since $x_k \in \mathbb{P}$ it
follows that $k < i$, thus $max(w_j) < i$. This contradicts the
conditions of the Theorem~\ref{ddom_al02}.\\
{\bf Case 4:}
If $P_5$ exists, then $min(w_j) \geq k$, where $k$ is the index of
$x_{k+1}$ in $P_1$. Since $x_{k+1} \in
\mathbb{S}$ it follows that $k > i$, thus $min(w_j) > i$. This
contradicts the conditions of the Theorem~\ref{ddom_al02}.
\begin{flushright}
$\Box$ \\
\end{flushright}

\end{document}